\documentclass[prd,nofootinbib,showpacs,showkeys,preprint]{revtex4}

\usepackage{amsmath}
\usepackage{amssymb}
\usepackage[dvips]{graphicx}
\usepackage{rotating}
\usepackage{color} 
\usepackage{multirow} 

\def \nbb {$\beta\beta_{0\nu}$ }

\newcommand{\Ds}{\Delta m^2_{\odot}}
\newcommand{\Da}{\Delta m^2_{\textsc{A}}}

\newcommand{\AddrAHEP}{
  {\it AHEP Group, Instituto de F\'{\i}sica Corpuscular --
    C.S.I.C./Universitat de Val{\`e}ncia \\
    Edificio de Institutos de Paterna, Apartado 22085,
  E--46071 Val{\`e}ncia, Spain}}

\newcommand{\AddrLisb}{%
 Departamento de F\'\i sica and CFTP, Instituto Superior T\'ecnico\\
          Av. Rovisco Pais 1, $\:\:$ 1049-001 Lisboa, Portugal }

\newcommand{\AddrWur}{%
Institut f\"ur Theoretische Physik und Astronomie, 
Universit\"at W\"urzburg\\
Am Hubland, 
97074 Wuerzburg}

\def\gsim{\raise0.3ex\hbox{$\;>$\kern-0.75em\raise-1.1ex\hbox{$\sim\;$}}}
\def\lsim{\raise0.3ex\hbox{$\;<$\kern-0.75em\raise-1.1ex\hbox{$\sim\;$}}}

\begin{document}

\preprint{IFIC/08-23}  

\title{Probing minimal supergravity in the type-I seesaw mechanism with lepton flavour
violation at the CERN LHC}


\author{M.~Hirsch} \email{mahirsch@ific.uv.es}\affiliation{\AddrAHEP}
\author{W. Porod} \email{porod@physik.uni-wuerzburg.de}\affiliation{\AddrWur}
\author{J.~C.~Romao}
\email{jorge.romao@ist.utl.pt}\affiliation{\AddrLisb}
\author{J.~W.~F.~Valle} \email{valle@ific.uv.es}
\affiliation{\AddrAHEP} \author{A.~Villanova del Moral}
\email{albert@cftp.ist.utl.pt}\affiliation{\AddrLisb}



\begin{abstract}
  The most general supersymmetric seesaw mechanism has too many
  parameters to be predictive and thus can not be excluded by {\em
    any} measurements of lepton flavour violating (LFV) processes. We
  focus on the simplest version of the type-I seesaw mechanism
  assuming minimal supergravity boundary conditions. We 
  compute branching ratios for the LFV scalar tau decays, ${\tilde
    \tau}_2 \to (e,\mu) + \chi^0_1$, as well as loop-induced LFV
  decays at low energy, such as $l_i \to l_j + \gamma$ and $l_i \to 3
  l_j$, exploring their sensitivity to the unknown seesaw parameters.
  We find some simple, extreme scenarios for the unknown right-handed
  parameters, where ratios of LFV branching ratios correlate with
  neutrino oscillation parameters. If the overall mass scale
  of the left neutrinos and the value of the reactor angle were known,
  the study of LFV allows, in principle, to extract information about
  the so far unknown right-handed neutrino parameters.
\end{abstract}

\maketitle

\section{Introduction}
\label{sec:int}

Neutrino oscillation experiments have demonstrated that neutrinos are
massive particles \cite{Fukuda:1998mi}. With the most recent
experimental data by the MINOS \cite{Collaboration:2007zza} and
KamLAND \cite{kl:2008ee} collaborations atmospheric and solar mass-squared differences are now known very precisely and global fits to
all neutrino oscillation data \cite{Maltoni:2004ei} also give quite
accurate determinations for the corresponding neutrino mixing
angles. For the overall mass scale of neutrinos and the third neutrino
mixing angle currently only upper limits exist, but considerable
progress is expected from future double beta decay
\cite{Avignone:2007fu} and reactor neutrino oscillation
\cite{Ardellier:2006mn,Guo:2007ug} experiments.

Neutrino masses provide the first experimental signal of physics
beyond the standard model (SM). From an experimental point of view,
neutrino oscillation data can easily be fitted in very much the
same way as the SM accounts for quark masses and mixings, i.e. namely
by Dirac neutrino masses.
From a theoretical point of view, however, such an ansatz is {\sl ad
  hoc} since, being electrically neutral, neutrinos are expected to be
Majorana particles~\cite{Schechter:1980gr}. Indeed, as noted already
in \cite{weinberg:1980mh}, the dimension-5 operator,
\begin{equation}\label{eq:dim5}
m_{\nu} = \frac{f}{\Lambda} (H L) (H L), 
\end{equation}
induces {\em Majorana} masses for neutrinos once the electroweak
symmetry breaks. 
This way the smallness of the neutrino masses can then be attributed
to the existence of some lepton number violating scale larger than the
electroweak scale.
A variety of ways to generate this operator have been suggested. The
resulting Majorana neutrino masses can be suppressed either by loop
factors, by a large mass scale, by a small scale whose absence
enhances the symmetry of theory, or by combinations of these
mechanisms~\cite{Altarelli:2004za}.

Electroweak scale models, such as, for example the Zee
model~\cite{Zee:1980ai}, the Babu-Zee model \cite{Babu:1988ki},
supersymmetric models with violation of R-parity
\cite{hall:1984id,Ross:1984yg,romao:1992ex,Hirsch:2000ef,Diaz:2003as}
or lepton number violating leptoquark models~\cite{Hirsch:1996qy}
generate neutrino masses at loop-level, resulting in $f\ll 1$ and
$\Lambda$ need not be much larger than $m_W$. A similar situation
arises in models like the inverse seesaw~\cite{Mohapatra:1986bd}. Such
low-scale models have the advantage that the new fields responsible
for the generation of neutrino masses may be directly accessible to
future accelerator experiments, see for example
\cite{AristizabalSierra:2006ri,AristizabalSierra:2006gb,Nebot:2007bc,%
AristizabalSierra:2007nf,hirsch:2002ys}.

The most popular mechanism to generate Majorana neutrino masses,
however, the celebrated seesaw mechanism~
\cite{Minkowski:1977sc,gell-mann:1980vs,yanagida:1979,mohapatra:1980ia,%
Schechter:1980gr,schechter:1982cv} 
assumes that lepton number is violated at a very large scale, probably
at energies comparable to the grand unification scale. This
``classical'' version of the seesaw mechanism, while automatically
suppressing neutrino masses without the need for any small pre-factor,
will unfortunately never be directly testable.

However {\em indirect} insight into the high-energy world might become
possible, if weak scale supersymmetry is realized in nature. Indeed,
starting from flavour diagonal soft supersymmetry (SUSY) breaking terms at some high-energy ``unification'' scale, flavour violation appears at lower
energies due to the renormalization group running of the soft breaking
parameters \cite{Hall:1985dx}.  If the (type-I) seesaw mechanism is
responsible for the observed neutrino masses, the neutrino Yukawa
couplings leave their imprint in the slepton mass matrices as shown
first in \cite{Borzumati:1986qx}.  Flavour off-diagonal entries in the
neutrino Yukawas then can lead to potentially large lepton flavour
violating lepton decays such as $l_i \to l_j +\gamma$ and $l_i \to 3
l_j$
\cite{Hisano:1995nq,Hisano:1995cp,Deppisch:2002vz,Arganda:2005ji,%
  Antusch:2006vw,Deppisch:2004fa} or $\mu-e$ conversion in nuclei
\cite{Arganda:2007jw,Deppisch:2005zm}.  In a similar spirit, if
supersymmetry is discovered at a future accelerator such as the LHC,
one can use measurements of masses and branching ratios of
supersymmetric particles to obtain indirect information on the range
of allowed seesaw parameters
\cite{Hisano:1998wn,Blair:2002pg,Freitas:2005et,Buckley:2006nv,Deppisch:2007xu}.
The most general supersymmetric seesaw mechanism has too many parameters 
to be predictive and thus can not be excluded by {\em any} measurements 
of lepton flavour violating (LFV) processes. Within the supersymmetric 
version of the seesaw
measurements of LFV observables outside the neutrino sector allow one
to obtain valuable independent information about the seesaw parameters
\cite{Ellis:2002fe}. There are two logical possibilities of how such
LFV measurements might be useful. (a) Given the current incomplete
knowledge on the light neutrino masses and angles, one could make some
simplified assumptions about the right-handed neutrino sector.  Then
``predictions'' for LFV observables as a function of the remaining
unknowns for the left-handed light neutrinos result. Or (b) one could
learn about the parameters of the right-handed neutrinos once the most
important, but currently unknown light neutrino observables have been
measured. While the second option might look more interesting, the
time scale for making progress on $m_{\nu}$, $s_{13}$ or the Dirac $CP$
phase $\delta$ will be long. Worse still, the Majorana phases of the
light neutrinos are unlikely to be ever reliably measured.
Hence experimental information most likely will be incomplete and
measurements of LFV observables will be useful to at least partially
reconstruct the seesaw parameters.

In this paper we study lepton flavour violating decays of the scalar
tau as well as LFV lepton decays at low energies. We assume minimal
supergravity (mSugra) boundary conditions and type-I seesaw as origin
of neutrino masses and mixings. We compare the sensitivities of
low-energy and accelerator measurements and study their dependence on
the most important unknown parameters. LFV measurements at
accelerators could be argued to be preferable to the low-energy LFV
experiments for ``reconstructing'' seesaw parameters, since from a
theoretical point of view they involve fewer assumptions. However, the
absolute values of LFV stau decays and, for example, Br($\mu \to e
+\gamma$) depend very differently on the unknown SUSY
spectrum. Whether low-energy LFV or LFV at accelerators yields more
insight into the seesaw mechanism can currently therefore not be
predicted.

While absolute values of LFV observables depend very strongly on the
soft SUSY breaking parameters, it turns out that ratios of LFV
branching ratios can be used to eliminate most of the dependence on
the unknown spectrum. I.e., ratios such as, for example,
Br(${\tilde\tau}_2 \to e + \chi^0_1$)/Br(${\tilde\tau}_2 \to \mu +
\chi^0_1$) are constants (for fixed neutrino parameters) over large
parts of the supersymmetric parameter space and therefore especially
suitable to extract information about the seesaw parameters. We
therefore study such ratios in detail, first in a useful analytical
approximation and then within a full numerical calculation.

The rest of this paper is organized as follows. In the next section, 
we will recall the basic features of the supersymmetric seesaw mechanism, 
mSugra and LFV in the slepton sector. Section \ref{sec:ana} then discusses 
analytical estimates for slepton mixing angles and the corresponding 
LFV observables. In Sec.~\ref{sec:num} we present our numerical 
results before concluding in Sec.~\ref{sec:cncl}.

\section{Setup: mSugra with  type I seesaw}
\label{sec:setup}

In order to fix the notation, let us briefly recall the main features
of the seesaw mechanism and mSugra. We will consider only the simplest
version of the seesaw mechanism here. It consists in extending the
particle content of the minimal supersymmetric standard model by three gauge singlet ``right-handed''
neutrino superfields. The leptonic part of the superpotential is thus
given by
\begin{equation}\label{su_pot}
W  =  Y_{e}^{ji}{\widehat L}_i{\widehat H_d}{\widehat E^c}_j
  + Y_{\nu}^{ji}{\widehat L}_i{\widehat H_u}{\widehat N^c}_j
  + M_{i}{\widehat N^c}_i{\widehat N^c}_i.
\end{equation}
where $Y_e$ and $Y_{\nu}$ denote the charged lepton and neutrino
Yukawa couplings, while ${\widehat N^c}_i$ are the ``right-handed''
neutrino superfields with $M_{i}$ Majorana mass terms of unspecified
origin. Since the ${\widehat N^c}_i$ are singlets, one can always
choose a basis in which the Majorana mass matrix of the right-handed
neutrinos is diagonal $\hat M_R$.

Note that LFV arises from supersymmetric as well as from gauge boson
loop diagrams, for example slepton-gaugino exchange loops and W loops
involving right-handed neutrino exchange. The former (SUSY-induced LFV) 
can be described by taking a basis where the $Y_e$ Yukawa coupling
matrix is diagonal, its entries fixed by the observed charged lepton
masses. This reduces the relevant physical parameters to a total of
21. 

While in extended schemes like inverse seesaw 
\cite{Bernabeu:1987gr,Deppisch:2004fa,Deppisch:2005zm} gauge-induced 
LFV is potentially sizeable, it is negligible in the simplest type-I 
seesaw model, due to the large values of $M_i$ required. Therefore, 
we focus on such intrinsically supersymmetric LFV, which can be 
well characterized Eq.~(\ref{su_pot}) in the unbroken SU(2) limit. 

Different parametrizations for the simplest seesaw have been discussed 
in the literature. The most convenient choice for our calculation is to 
go to the basis
where the charged lepton mass matrix is diagonal. We then have as
parameters 9 mass eigenstates (3 charged leptons, the 3 light and the
3 heavy neutrinos). The remaining 12 parameters can be encoded in two
matrices $V_L$ and $V_R$, with 3 angles and 3 phases each, which
diagonalize $Y_{\nu}$,
\begin{equation}\label{diagYnu}
{\hat Y_{\nu}}= V_R^{\dagger} Y_{\nu} V_L.
\end{equation}
The effective mass matrix of the left-handed neutrinos is given 
in the usual seesaw approximation as
\begin{equation}\label{meff}
m_{\nu} = - \frac{v_U^2}{2} Y_{\nu}^T\cdot M_{R}^{-1}\cdot Y_{\nu}.
\end{equation}
If one of the $M_{i}$ eigenvalues of the matrix $M_R$ goes to infinity
(or the corresponding vector in $Y_{\nu}$ to zero) the corresponding
eigenvalue of $m_{\nu}$ ($m_i$) goes to zero. Since the neutrino mass
matrix is complex symmetric, Eq.~(\ref{meff}) is diagonalized
by~\cite{Schechter:1980gr}
\begin{equation}\label{diagmeff}
{\hat m_{\nu}} = U^T \cdot m_{\nu} \cdot U.
\end{equation}
Inverting the seesaw equation, Eq.~(\ref{meff}), 
allows one to express $Y_{\nu}$ as \cite{Casas:2001sr}
\begin{equation}\label{Ynu}
Y_{\nu} =\sqrt{2}\frac{i}{v_U}\sqrt{\hat M_R}\cdot R\cdot\sqrt{{\hat m_{\nu}}}\cdot U^{\dagger}.
\end{equation}
where $\hat m_{\nu}$ is the diagonal matrix with $m_i$ eigenvalues and
$R$ in general is a complex orthogonal matrix. Note, that in the
special case $R=1$, $Y_{\nu}$ contains only ``diagonal'' products 
$\sqrt{M_im_{i}}$. Note that in this approximation the 18 parameters in 
$Y_{\nu}$ are reduced to 12, which are expressed as six light neutrino 
mixing angles and phases in the lepton mixing matrix $U$, the 3 light 
neutrino masses in $\hat m_{\nu}$ and the 3 heavy `` right-handed'' 
neutrino masses in $\sqrt{\hat M_R}$. 

In the general MSSM, LFV off-diagonal entries in the slepton mass
matrices are free parameters. In order to correlate LFV in the slepton
sector with the LFV encoded in $Y_{\nu}$ one must assume some scheme
for supersymmetry breaking. We will restrict ourselves here to the
case of mSugra, characterized by four continuous and one discrete free
parameter,  usually denoted as
\begin{equation}\label{sugra-par}
m_0, \hskip2mm M_{1/2}, \hskip2mm A_0, \hskip2mm \tan\beta, \hskip2mm 
{\rm Sgn}(\mu)
\end{equation}
Here, $m_0$ is the common scalar mass, $M_{1/2}$ the gaugino mass 
and $A_0$ the common trilinear parameter, all defined at the 
grand unification scale, $M_{X} \simeq 2 \cdot 10^{16}$ GeV. 
The remaining two parameters are $\tan\beta = v_U/v_D$ and the 
sign of the Higgs mixing parameter $\mu$. 
For reviews on mSugra, see, for example 
\cite{Haber:1984rc,Martin:1997ns}.

Calculable LFV entries appear in the slepton mass matrices, due to the
nontrivial generation structure of the neutrino Yukawa matrix in
Eq.~(\ref{su_pot}), as first pointed out in \cite{Borzumati:1986qx}.
In order to determine their magnitude we solve the complete set of
renormalization group equations, given
in~\cite{Hisano:1995cp,Antusch:2002ek}. 
It is however useful for a qualitative understanding, to
consider first the simple solutions to the renormalization group equations found in the leading
log approximation \cite{Hisano:1995cp}, given by
\begin{eqnarray}\label{running}
(\Delta M_{\tilde L}^2)_{ij} =
 -\frac{1}{8\pi^2}(3 m_0^2 + A_0^2) 
  (Y_{\nu}^{\dagger}LY_{\nu})_{ij} \\ \nonumber
(\Delta A_l)_{ij} =  -\frac{3}{8\pi^2}A_0Y_{l_i}
   (Y_{\nu}^{\dagger}LY_{\nu})_{ij}\\ \nonumber
(\Delta M_{\tilde E}^2)_{ij} = 0,
\end{eqnarray}
where only the parts proportional to the neutrino Yukawa couplings
have been written. The factor $L$ is defined as
\begin{equation}\label{deffacL}
L_{kl} = \log\Big(\frac{M_X}{M_{k}}\Big)\delta_{kl}. 
\end{equation}
Equation~(\ref{running}) shows that, within the type-I seesaw mechanism 
the right slepton
parameters do not run in the leading-log approximation.  Thus, LFV
scalar decays should be restricted to the sector of left-sleptons in
practice, apart from left-right mixing effects which could show up in
the scalar tau sector. Also note that for the trilinear parameters
running is suppressed by charged lepton masses.

Note also that the LFV slepton mass-squareds involve a different 
combination of neutrino Yukawas 
and right-handed neutrino masses than the left-handed neutrino masses 
of Eq.~(\ref{meff}). In fact, since $(Y_{\nu}^{\dagger}LY_{\nu})$ is
a hermitian matrix, it obviously contains only nine free parameters~\cite{Ellis:2002fe}, the same number of unknowns as on the right-hand
side of Eq.~(\ref{Ynu}), given that in principle all 3 light neutrino
masses, 3 mixing angles and 3 $CP$ phases are potentially measurable
\footnote{In practice measuring the
  unknown angle $\theta_{13}$ and the Dirac $CP$ phase requires improved
  neutrino oscillation studies~\cite{Nunokawa:2007qh} and will not be
  an easy task. Even if we are lucky to measure the overall neutrino
  mass scale in \nbb experiments~\cite{Avignone:2007fu}, the Majorana
  phases contained in $U$ are much harder to determine in practice.}.

In an ideal world where all low energy paramaters, namely the 3 light
neutrino masses, 3 mixings and 3 $CP$ violation parameters were known,
the remaining parameters entering Eq.~(\ref{su_pot}) could in
principle be reconstructed by measuring all entries in $(\Delta
M_{\tilde L}^2)_{ij}$. This would determine the full set of 18+3
parameters which, to a good approximation, characterize LFV in the
minimal type-I seesaw.
In practice, however, there are two obstacles. (i) Calculability of
$(\Delta M_{\tilde L}^2)_{ij}$ using Eq.~(\ref{running}) assumes
implicitly that there are no threshold effects near the unification
scale which destroy the strict proportionality to the parameters $m_0$
and $A_0$ \cite{Davidson:2004wi}.  In realistic grand unified theory models this might
not be the case. And (ii) it is not realistic to assume that all
entries in $(\Delta M_{\tilde L}^2)_{ij}$ can be measured with
sufficient accuracy, since (a) the diagonal shifts $(\Delta M_{\tilde
  L}^2)_{ii}$ are very small compared to $(M_{\tilde L}^2)_{ii}$
(nearly everywhere in the available parameter space) and (b) the
determination of the phases require to measure $CP$-violating LFV
observables. The latter does not seem to be a very realistic option
either, since, as our numerical results show, one expects only rather
low statistics to be available in measurements of LFV slepton decays.

\section{Analytical results for flavour violating processes}
\label{sec:ana}

In this section we present some general formulas describing lepton
flavour violation within type-I seesaw schemes. We concentrate on the
discussion of ratios of LFV branching ratios, since, as mentioned in
the introduction, these are most easily connected to the seesaw
parameters. As a first approximation we adopt the mass insertion
approximation, neglecting left-right mixing in the slepton mass matrix
and taking the leading-logs (see below).  We will demonstrate the
reliability of our analytical estimates in the next section, where we
perform a full numerical calculation of the various LFV branching
ratios, which does not rely on any of the approximations discussed in
this section.

\subsection{General formulas}

The charged slepton mass matrix is a (6,6) matrix, containing left and
right sleptons. Here we concentrate exclusively on the left-slepton
sector.  Taking into account the discussion given in
Sec.~\ref{sec:setup}, this is a reasonable first approximation, as can
be seen from Eq.~(\ref{running}).
The left-slepton mass matrix is diagonalized by a matrix $R^{\tilde
  l}$, which in general can be written as a product of three Euler
rotations.  However, if the mixing between the different flavour
eigenstates is sufficiently small, $R^{\tilde l}$ can be approximated
as
\begin{equation}\label{defR}
R^{\tilde l} \simeq
\left(\begin{array}{cccc}
1 & \theta_{{\tilde e}{\tilde\mu}} & \theta_{{\tilde e}{\tilde\tau}} \cr
- \theta_{{\tilde e}{\tilde\mu}} & 1 & \theta_{{\tilde \mu}{\tilde\tau}} \cr
- \theta_{{\tilde e}{\tilde\tau}}  & - 
\theta_{{\tilde \mu}{\tilde\tau}} & 1 \cr
\end{array}\right),
\end{equation}
an approximation that corresponds to that employed in the
mass-insertion method~\cite{Borzumati:1986qx}.
In this small-angle approximation each angle can be estimated by 
the following simple formula
\begin{equation}\label{slepAng}
\theta_{ij} \simeq \frac{(\Delta M_{\tilde L}^2)_{ij}}
       {(\Delta M_{\tilde L}^2)_{ii} - (\Delta M_{\tilde L}^2)_{jj}}.
\end{equation}
LFV decays are directly proportional to the squares of these mixing 
angles, for example 
Br$(\mu \to e +\gamma) \sim (\theta_{{\tilde e}{\tilde\mu}})^2$ 
if all angles are small. 

Within mSugra ratios of LFV branching ratios can then be used to 
minimize the dependence of observables on SUSY parameters. Consider 
the case of LFV decays which involve only one generation of sleptons, 
for example Br(${\tilde\tau}_2 \to  e + \chi^0_1$) and 
Br(${\tilde\tau}_2 \to  \mu + \chi^0_1$). Taking the ratio of 
these two decays
\begin{equation}\label{mainresult}
\frac{Br({\tilde\tau}_2 \to  e +\chi^0_1)}
     {Br({\tilde\tau}_2 \to  \mu +\chi^0_1)}
 \simeq \Big(\frac{\theta_{{\tilde e}{\tilde\tau}}}
{\theta_{{\tilde\mu}{\tilde\tau}}}\Big)^2 
 \simeq 
\Big(\frac{(\Delta M_{\tilde L}^2)_{13}}{(\Delta M_{\tilde L}^2)_{23}}\Big)^2,
\end{equation}
i.e. one expects that (a) all the unknown SUSY mass parameters and (b)
the denominators of Eq.~(\ref{slepAng}) cancel approximately.  The
latter should happen practically everywhere in mSugra parameter space
since $(M_{\tilde L}^2)_{ee}\simeq (M_{\tilde L}^2)_{\mu\mu}$.  This
straightforward observation forms the basis for our claim that {\em
ratios of branching ratios} are the theoretically cleanest way to
learn about the unknown seesaw parameters. Numerically we have found,
that relations similar to Eq.~(\ref{mainresult}) hold also for ratios
of observables involving decaying particles of different generations, 
such as the low-energy ratio Br$(\mu \to e +\gamma)$/Br$(\tau \to e +\gamma)$. 

To calculate estimates for the different ratios of branching ratios 
we therefore define
\begin{equation}
\label{eq:def-rijkkl}
r^{ij}_{kl}\equiv \frac{|(\Delta M_{\tilde{L}}^2)_{ij}|}
                      {|(\Delta M_{\tilde{L}}^2)_{kl}|}
\end{equation}
where the observable quantity is $(r^{ij}_{kl})^2$. Of course, only
two of the three possible combinations that can be formed are
independent.  For example, Br$(\mu \to e +\gamma)$/Br$(\tau \to e
+\gamma) \simeq (r^{12}_{13})^2\times{\cal R}$. Here, ${\cal R}$ is a
correction factor taking into account the different total widths of
the muon and the tau, ${\cal
  R}=\Gamma_{\tau}/\Gamma_{\mu}$~\footnote{The inclusion of this
  factor (and similar corrections for the other low-energy LFV decays)
  is necessary, since $(r^{ij}_{kl})^2$ relate really partial widths,
  whereas the measured quantity is usually the branching ratio.}.

In the leading-log approximation the off-diagonal elements of the 
charged slepton mass matrix are proportional to 
$(\Delta M_L^2)_{ij}\propto 
\left((Y^{\nu})^{\dagger}L
(Y^{\nu})\right)_{ij}$. 
Using the parametrization for the Yukawa couplings of Eq.~(\ref{Ynu})
the entries in $(\Delta M_L^2)_{ij}$ can be expressed as
\begin{equation}\label{wq:delmgen}
(\Delta {M_L^2})_{ij} 
\propto U_{i\alpha}U_{j\beta}^*\sqrt{m_{\alpha}}\sqrt{m_{\beta}}
R_{k\alpha}^*R_{k\beta}M_k\log\left(\frac{M_X}{M_k}\right).
\end{equation}
We can now rewrite Eq.~(\ref{wq:delmgen}) in terms of observables
which are more directly related to experiments. In the standard
parametrization for the leptonic mixing matrix $U$ is completely
analogous to the CKM matrix and can be written as
\begin{equation}
U=
\begin{pmatrix}
c_{12} c_{13} & s_{12} c_{13} & s_{13} e^{-i\delta} \\
-s_{12} c_{23} - c_{12} s_{23} s_{13} e^{i\delta} & 
c_{12} c_{23} - s_{12} s_{23} s_{13} e^{i\delta} & s_{23} c_{13} \\
s_{12} s_{23} - c_{12} c_{23} s_{13} e^{i\delta} & 
- c_{12} s_{23} - s_{12} c_{23} s_{13} e^{i\delta} & c_{23} c_{13} 
\end{pmatrix}
\end{equation}
where we assumed strict unitarity and neglected the Majorana
phases~\cite{Schechter:1980gr}, because they do not affect lepton
number conserving processes such as the LFV decays we are concerned
with here.

Given that neutrino oscillation experiments fix two mass-squared
splittings, we can re-express the three light neutrino masses in terms
of one overall neutrino mass scale and the measured quantities $\Ds$
and $\Da$, where $\Ds$ ($\Da$) is the solar (atmospheric) mass-squared
splitting. We will refer to the case of $m_1\equiv 0$ ($m_3\equiv 0$)
as strict normal (inverse) hierarchy. This choice has the advantage
that in both cases $s_{12}\equiv \sin\theta_{\odot}$ and $s_{23}\equiv
\sin\theta_{\textsc{a}}$. Equation~(\ref{wq:delmgen}) can then be written
in terms of the measured neutrino angles $s_{12}$ and $s_{23}$, the
measured neutrino mass-squared splittings, plus the so far unknown
overall neutrino mass scale $m_{\nu}$ and the reactor neutrino angle
$s_{13}\equiv s_{R}$. If the latter were measured, one could extract
information on the right-handed neutrino mass scale and/or the matrix
$R$ from Eq.~(\ref{wq:delmgen}). Conversely, we could learn about
$m_{\nu}$ and $s_{13}$ from measurements of LFV decays, making some
assumptions about the scale $M_R$ and the possible textures of the
Yukawa couplings that determine  $M_R$ and $R$.

\subsection{Degenerate right-handed neutrinos}

In this subsection we will assume that the three right-handed
neutrinos are degenerate. This simplifying ansatz allows us to study
the sensitivity with a single mass-scale parameter associated with the
neutrino mass generation via type-I seesaw mechanism. This ansatz can
be theoretically motivated in the framework of some flavour
symmetries, for example $A_4$~\cite{Hirsch:2008rp}. In the special
case that the matrix $R$ is real, Eq.~(\ref{wq:delmgen}) reduces
to
\begin{eqnarray}
\label{delmR1}
\left(\Delta{M_{\tilde{L}}^2}\right)_{12} & 
\propto & c_{12} c_{13} \left( - s_{12}c_{23} - c_{12}s_{23}s_{13} 
e^{-i\delta} \right) z_1 \\ \nonumber
& + & s_{12}c_{13} \left(c_{12}c_{23} - s_{12}s_{23}s_{13} 
e^{-i\delta} \right) z_2 + s_{23}c_{13}s_{13} e^{-i\delta}  z_3  \\ \nonumber
\left(\Delta{M_{\tilde{L}}^2}\right)_{13} & 
\propto & c_{12} c_{13} \left(s_{12} s_{23} - c_{12}c_{23}s_{13} 
e^{-i\delta} \right) z_1 \\ \nonumber
& + & s_{12}c_{13} \left( - c_{12}s_{23} - s_{12} c_{23}s_{13} 
e^{-i\delta} \right) z_2 + c_{23}c_{13}s_{13} e^{-i\delta} z_3  \\ \nonumber
\left(\Delta{M_{\tilde{L}}^2}\right)_{23} & 
\propto  & \left( s_{12}s_{23} - c_{12}c_{23}s_{13} 
e^{-i\delta} \right) \left(-s_{12}c_{23} - c_{12}s_{23}s_{13} 
e^{i\delta} \right) z_1 \\ \nonumber
& + & \left( - c_{12}s_{23} - s_{12}c_{23} s_{13} 
e^{-i\delta} \right) \left( c_{12}c_{23} - s_{12}s_{23}s_{13} 
e^{i\delta} \right) z_2 \\ \nonumber
& + & s_{23}c_{23}c_{13}^2 z_3,
\end{eqnarray}
where 
\begin{equation}
z_i\equiv m_i M_i \log\left(\frac{M_X}{M_i}\right).
\end{equation}
For this degenerate right-handed neutrino ansatz the combination
$M_i\log (\frac{M_X}{M_i})$ becomes an overall factor, which can be
taken out from Eq.~(\ref{wq:delmgen}), since it cancels upon taking
ratios. I.e. for degenerate right-handed neutrinos one may simply make
the replacement $z_i \to m_{i}$ in Eq.~(\ref{delmR1}).

As a starting approximation for the following estimates, let us assume
that the lepton mixing matrix has the exact tribimaximal (TBM)
form~\cite{Harrison:2002er}
\begin{equation}\label{tbm}
{U}={U}_{\textrm{TBM}}=
\begin{pmatrix}
\sqrt{\frac{2}{3}} & \frac{1}{\sqrt{3}} & 0 \\
-\frac{1}{\sqrt{6}} & \frac{1}{\sqrt{3}} & \frac{1}{\sqrt{2}} \\
 \frac{1}{\sqrt{6}} & -\frac{1}{\sqrt{3}} & \frac{1}{\sqrt{2}} 
\end{pmatrix} .
\end{equation}
As is well-known, Eq.~(\ref{tbm}) is an excellent first-order approximation 
to the measured neutrino mixing angles \cite{Maltoni:2004ei}. With 
this assumption the ratios of the off-diagonal elements of the charged 
slepton mass matrix are simply given by 
\begin{eqnarray}\label{tbm_1}
r^{12}_{13}  & =  & 1 \\ \nonumber
r^{12}_{23}  & = & r^{13}_{23}  = \frac{2 (m_2-m_1)}{|3 m_3-2 m_2-m_1|}.
\end{eqnarray}
As Eq.~(\ref{tbm_1}) shows, $r^{12}_{23}$ and $r^{13}_{23}$ depend on mass 
squared splittings and on the overall neutrino mass scale, i.e. also on the 
unknown neutrino mass hierarchy. In the case of strict normal hierarchy 
(SNH, $m_{1}\equiv 0$)
\begin{equation}\label{tbmSNH}
r^{12}_{23}  = r^{13}_{23} = 
\frac{2\sqrt{\alpha}}{3\sqrt{1+\alpha}-2\sqrt{\alpha}}
\end{equation}
where $\alpha \equiv \frac{\Ds}{|\Da|}$, while for the case of strict 
inverse hierarchy (SIH, $m_{3}\equiv 0$) 
\begin{equation}
r^{12}_{23}  = r^{13}_{23} = \frac{2(1-\sqrt{1-\alpha})}{2+\sqrt{1-\alpha}}.
\end{equation}
Finally, for quasi-degenerate (QD) neutrinos, defined as
$\sqrt{\Da}\ll m_{\nu}$, one finds
\begin{equation}
r^{12}_{23}  = r^{13}_{23} \simeq \frac{2\alpha}{3\sigma_{\textsc{a}}+\alpha}
\end{equation}
where $\sigma_{\textsc{a}}$ is the sign of the atmospheric mass splitting
\begin{equation}
\sigma_{\textsc{a}}\equiv\frac{\Da}{|\Da|}.
\end{equation}
Note that $\sigma_{\textsc{a}}$ equals $+1$ ($-1$) for normal
(inverse) hierarchy. Thus QD neutrinos with normal (QDNH) or inverse
hierarchy (QDIH) lead formally to different results. However, this
difference is numerically not relevant, once uncertainties are taken
into account.

\begin{figure}[htb] \centering
\includegraphics[height=60mm,width=80mm]
{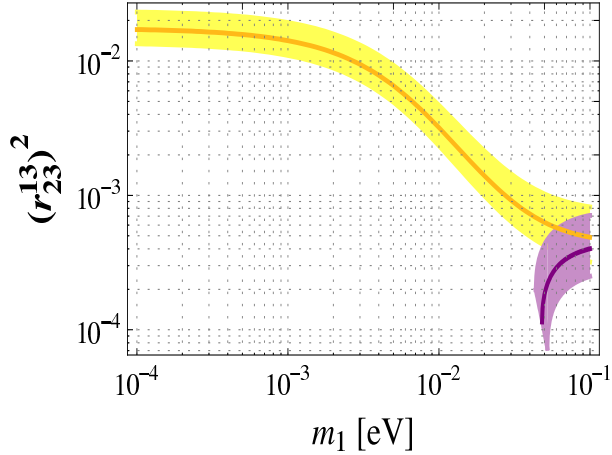}
\caption{Ratio $(r^{13}_{23})^2$ versus the neutrino mass $m_1$ 
in eV. The light/yellow (dark/violet) band is for the case of normal 
(inverse) hierarchy. The width of the band indicates the uncertainty 
due to the currently allowed 3 $\sigma$ C.L. ranges for $\Da$ and 
$\Ds$. The calculation assumes exact tribimaximal mixing for the 
left-handed neutrinos.}
\label{fig:degana}
\end{figure}

Figure~\ref{fig:degana} shows the ratio $(r^{13}_{23})^2$ versus the
neutrino mass $m_1$ in eV for normal (inverse) hierarchy. The figure
demonstrates the importance of the absolute neutrino mass scale for
$(r^{13}_{23})^2$. In the most general case one must use
Eqs.~(\ref{eq:def-rijkkl}) and~(\ref{delmR1}). However, for $s_{13}=0$
the explicit dependence of the ratios of the off-diagonal elements of
the charged slepton mass-squared matrix on the other neutrino angles
matrix follow rather simple expressions
\begin{eqnarray}
r^{12}_{13} & = &\frac{c_{23}}{s_{23}},\\ \nonumber 
r^{12}_{23} & = & \frac{1}{s_{23}}s_{12}c_{12}
\frac{m_2-m_1}{|m_3-c_{12}^2m_2-s_{12}^2m_1|},\\ \nonumber 
r^{13}_{23} & = & \frac{1}{c_{23}}s_{12}c_{12}
\frac{m_2-m_1}{|m_3-c_{12}^2m_2-s_{12}^2m_1|}.
\end{eqnarray}

\begin{figure}[htb] \centering
\includegraphics[height=50mm,width=80mm]{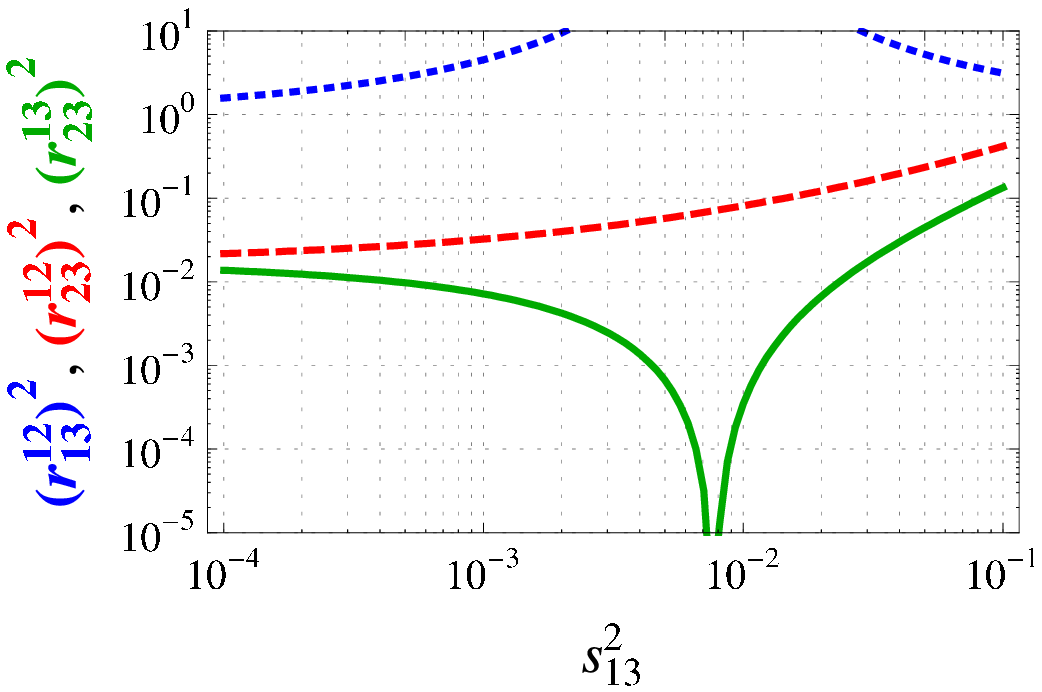}
\includegraphics[height=50mm,width=80mm]{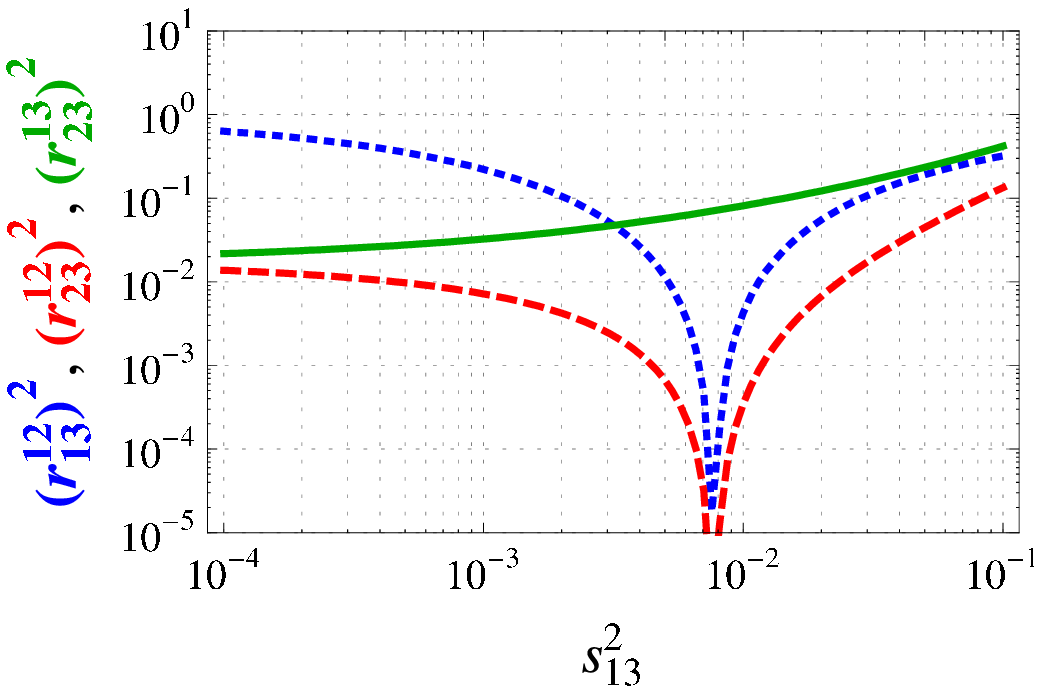}\\
\includegraphics[height=50mm,width=80mm]{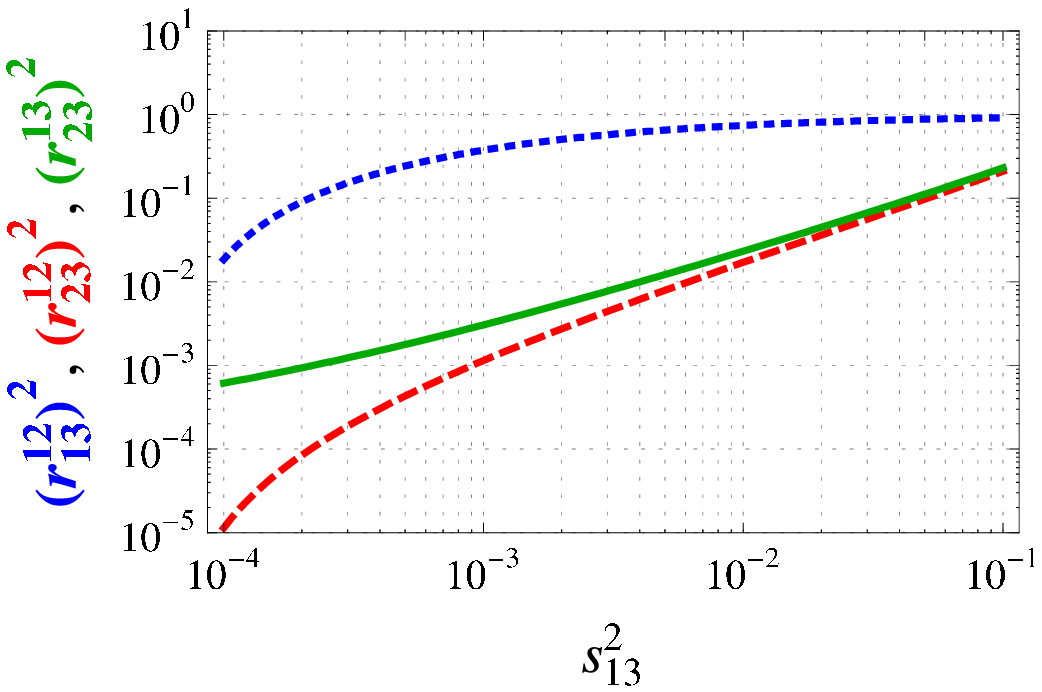}
\includegraphics[height=50mm,width=80mm]{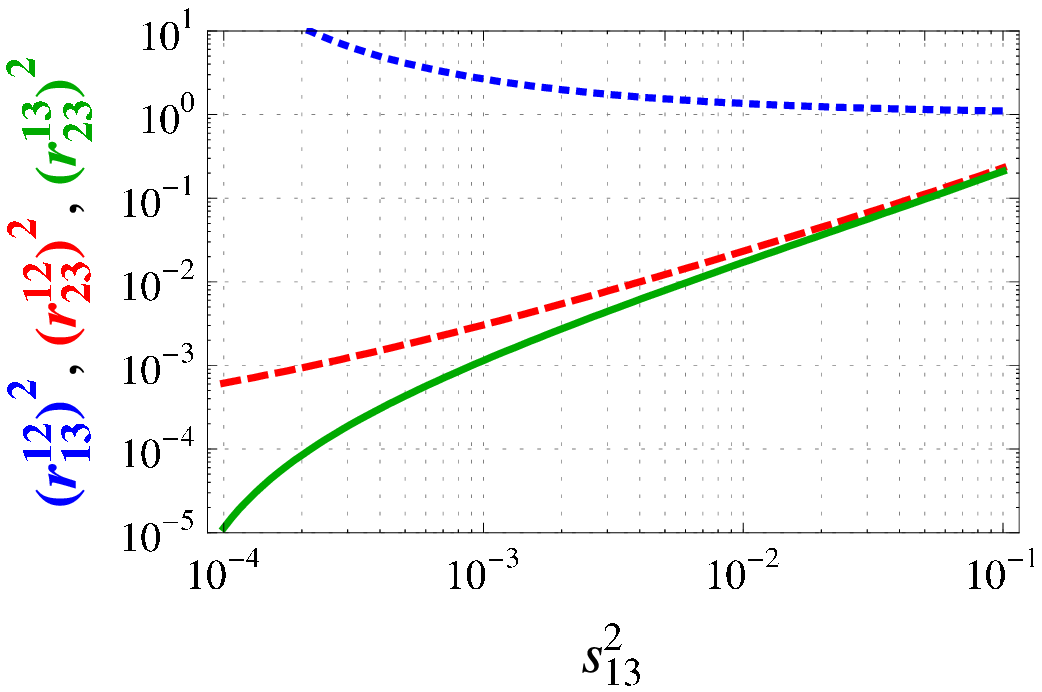}\\
\vskip-5mm
\caption{\label{fig:r-vs-s13-degM}Square ratios $(r^{12}_{13})^2$ (blue line, 
dotted line), $(r^{12}_{23})^2$ (red line, dashed line) and $(r^{13}_{23})^2$ 
(green line, full line) versus $s_{13}^2$ for SNH (upper panels), SIH 
(lower panels) for $\delta=0$ (left panels) and 
$\delta=\pi$ (right panels). The plots assume that the heavy neutrinos 
are degenerate. The other light neutrino parameters have been fixed to 
their b.f.p. values. Note from Eq.~(\ref{delmR1}), that for $\tan^2\theta_{\rm A}=1$, 
$(r^{12}_{23})^2$ and $(r^{13}_{23})^2$ are symmetric under the exchange 
of $\delta=0$ $\leftrightarrow$ $\delta=\pi$.}
\end{figure}

\begin{figure}[htb] \centering
\includegraphics[height=50mm,width=80mm]{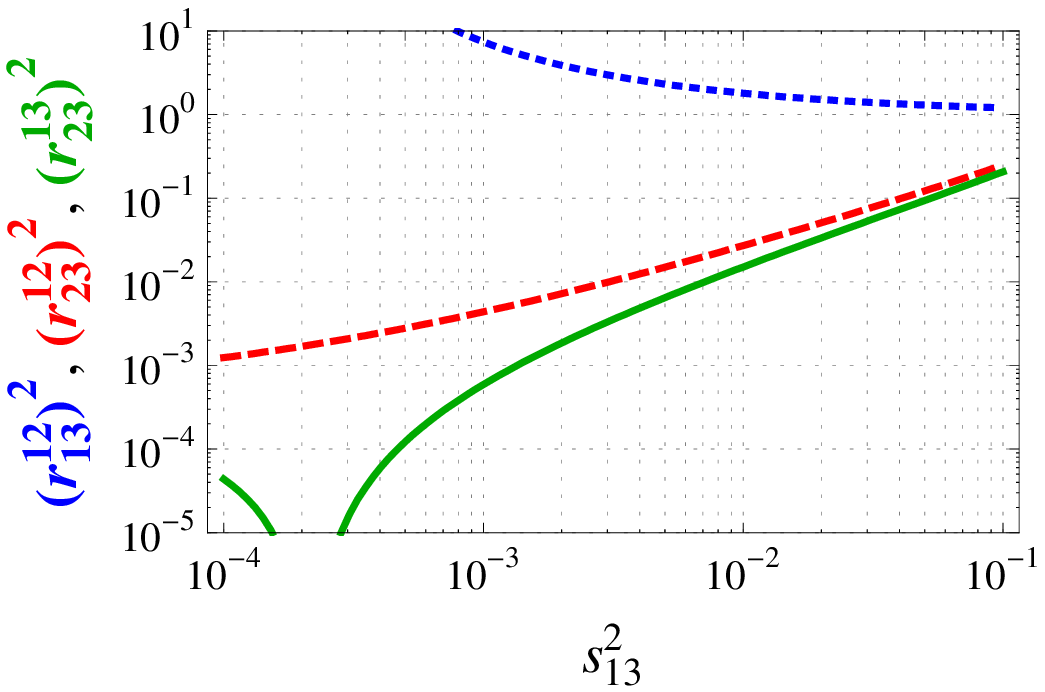}
\includegraphics[height=50mm,width=80mm]{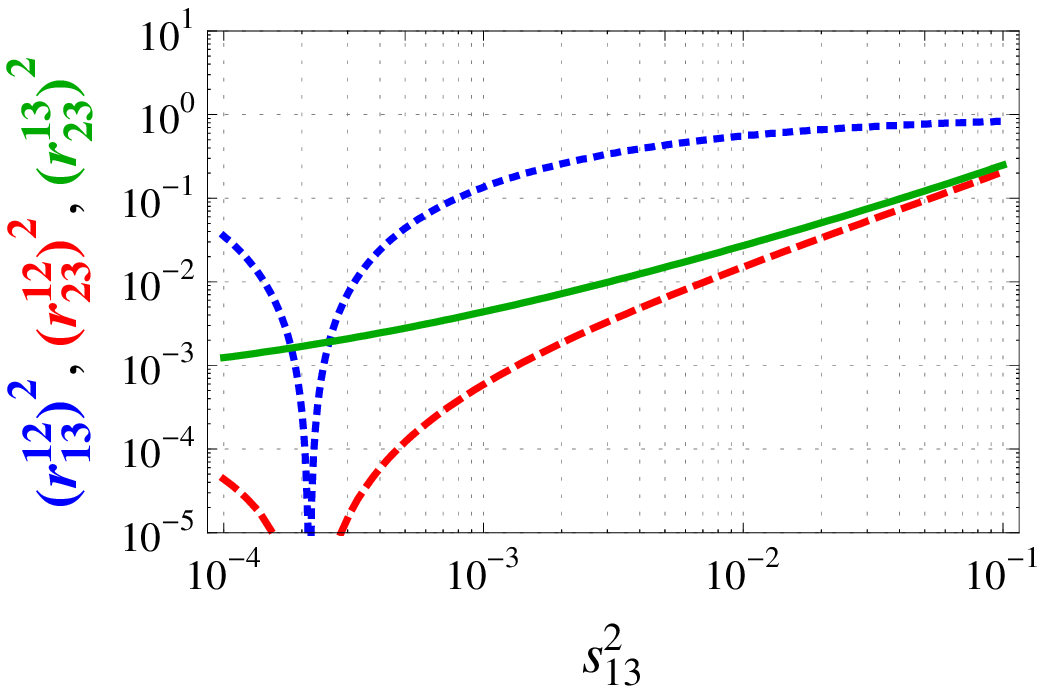}\\
\includegraphics[height=50mm,width=80mm]{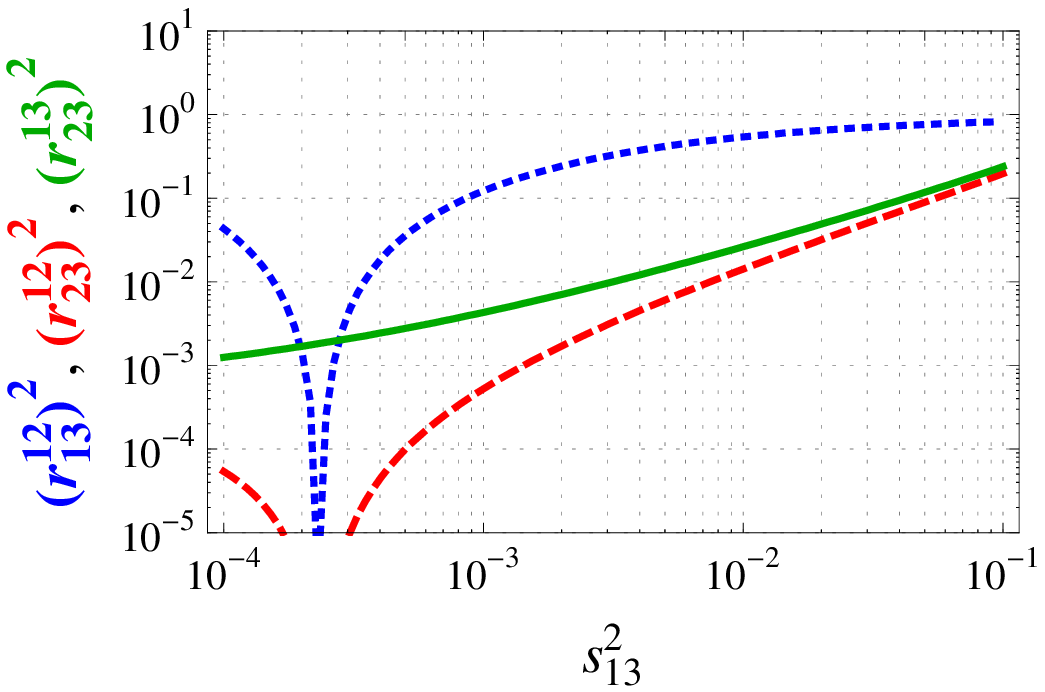}
\includegraphics[height=50mm,width=80mm]{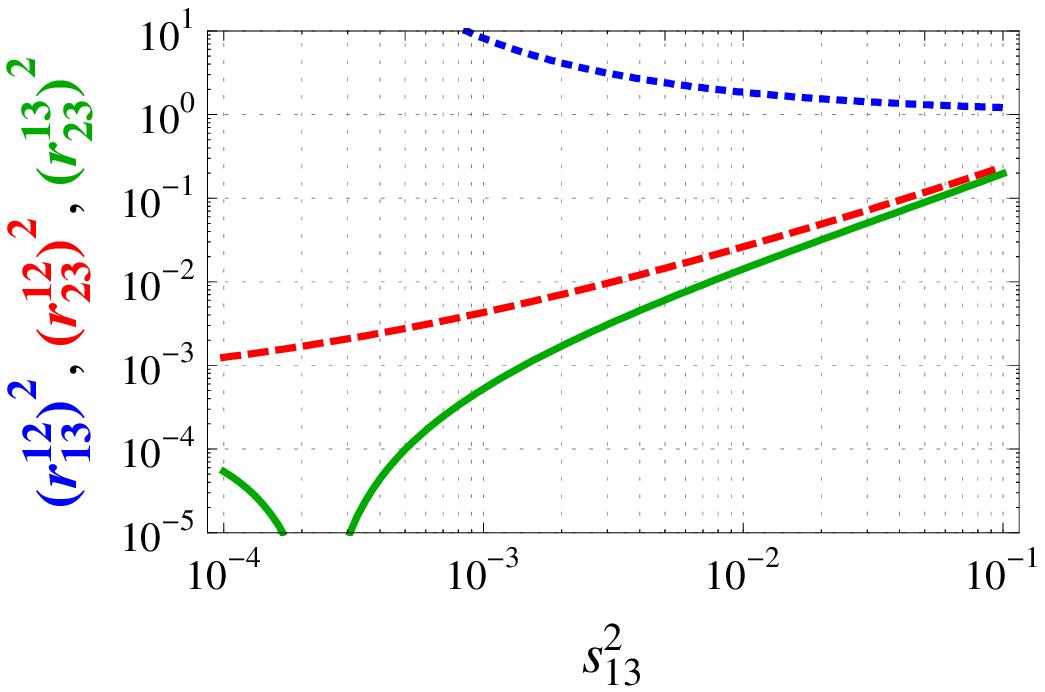}\\
\vskip-5mm
\caption{\label{fig:r-vs-s13-degM_degNu} As
  Fig.~\ref{fig:r-vs-s13-degM}, but for the limit of
  quasi-degenerate light neutrinos. Square ratios $(r^{12}_{13})^2$
  (blue line, dotted line), $(r^{12}_{23})^2$ (red line, dashed line)
  and $(r^{13}_{23})^2$ (green line, full line) versus $s_{13}^2$ for
  QDNH (upper panels), QDIH (lower panels) for $\delta=0$ (left
  panels) and $\delta=\pi$ (right panels).}
\end{figure}

Figures~\ref{fig:r-vs-s13-degM} and \ref{fig:r-vs-s13-degM_degNu} show
the dependence of the square ratios $(r^{ij}_{kl})^2$ as a function of
$s_{13}^2$ for the different extreme cases of SNH and SIH as well as
QDNH and QDIH, for two choices of the Dirac phase $\delta=0,\,\pi$.
These ratios $(r^{ij}_{kl})^2$ depend strongly on $s_{13}^2$. Note
from Eq.~(\ref{delmR1}) that for $\tan^2\theta_{\rm A}=1$,
$(r^{12}_{23})^2$ and $(r^{13}_{23})^2$ are invariant under exchange
of $\delta=0$ $\leftrightarrow$ $\delta=\pi$. If $\tan^2\theta_{\rm
  A}\ne 1$, this symmetry is broken, but always one of the two ratios
$r^{12}_{23}$ and $r^{13}_{23}$ is guaranteed to be non-vanishing
regardless of the value of $s_{13}$. A non-zero measurement of both
ratios would therefore in principle contain information on both
$s_{13}$ and $\delta$ (if right-handed neutrinos are degenerate).

Figure \ref{fig:r-vs-s13-degM_degNu} shows also that the cases QDNH
and QDIH are also symmetric under the simultaneous exchange of
$\delta=0$ $\leftrightarrow$ $\delta =\pi$ and QDNH $\leftrightarrow$
QDIH, for the case of $\tan^2\theta_{\rm A}=1$. This symmetry is
broken in all cases for $\tan^2\theta_{\rm A}\ne 1$, as seen from the
numerical values given in Tables~\ref{tab:NHandIH} and
\ref{tab:QD}. Tables~\ref{tab:NHandIH} and \ref{tab:QD} show
numerical values for $r^{ij}_{kl}$ for the extreme cases of SNH, SIH,
QDNH and QDIH for various different choices of neutrino parameters.
In the rows labeled as TBM, we have used the TBM values for
$\theta_{12}$ and $\theta_{23}$ and the neutrino mass splittings have
been fixed to their best-fit values taken from
Ref.~\cite{Maltoni:2004ei}. In the rows labeled as $3\sigma$, we take
into account the experimentally allowed $3\sigma$ ranges for neutrino
oscillation parameters: $s_{12}^2=0.26-0.40$, $s_{23}^2=0.34-0.67$,
$\Ds =(7.1-8.3)\times 10^{-5} {\rm eV^2}$ and $\Da=(2.0-2.8)\times
10^{-3} {\rm eV^2}$.  In the first column, $\theta_{13}$ has been
fixed to its TBM value ($s_{13}=0$), while in the second and third
columns $s_{13}$ has been fixed to $s_{13}^{\rm max}$, which is the
experimentally allowed maximum value: $(s_{13}^{\rm max})^2=0.050$ at
$3\sigma$ C.L. In the second column, the Dirac phase is fixed to
$\delta=0$, while in the third column $\delta=\pi$. Note that, as
already mentioned, these estimates are valid in the small mixing
limit and hence these values are indicative only.
\begin{table}[htb]\begin{center}
\begin{tabular}{|c|c|c||r@{.}l|r@{.}l|r@{.}l|}\hline
\multicolumn{3}{|c||}{} & \multicolumn{2}{|c|}{$s_{13}=0$} & \multicolumn{2}{|c|}{$s_{13}=s_{13}^{\rm max},\, \delta=0$} & \multicolumn{2}{|c|}{$s_{13}=s_{13}^{\rm max},\, \delta=\pi$} \\\hline\hline
\multirow{6}{*}{SNH} 
& \multirow{3}{*}{TBM}
 & $(r^{12}_{13})^2$  & $1$&$0$              & $5$&$2$               & $1$&$9\times 10^{-1}$ \\
 && $(r^{12}_{23})^2$ & $1$&$7\times 10^{-2}$ & $2$&$3\times 10^{-1}$ & $4$&$4\times 10^{-2}$ \\
 && $(r^{13}_{23})^2$ & $1$&$7\times 10^{-2}$ & $4$&$4\times 10^{-2}$ & $2$&$3\times 10^{-1}$ \\\cline{2-9}
& \multirow{3}{*}{$3\sigma$}
 & $(r^{12}_{13})^2$  & $[0$&$49,\, 1.9]$              & $[1$&$8,\, 35]$                & $[0$&$33,\, 5.7]\times 10^{-1}$ \\
 && $(r^{12}_{23})^2$ & $[0$&$91,\, 3.6]\times 10^{-2}$ & $[2$&$0,\, 3.2]\times 10^{-1}$ & $[0$&$96,\, 12] \times 10^{-2}$ \\
 && $(r^{13}_{23})^2$ & $[0$&$92,\, 3.7]\times 10^{-2}$ & $[0$&$87,\, 11]\times 10^{-2}$ & $[2$&$0,\, 3.2] \times 10^{-1}$ \\\hline
\multirow{6}{*}{SIH} 
& \multirow{3}{*}{TBM}
 & $(r^{12}_{13})^2$  & $1$&$0$               & $8$&$7\times 10^{-1}$ & $1$&$1$ \\
 && $(r^{12}_{23})^2$ & $1$&$1\times 10^{-4}$ & $9$&$7\times 10^{-2}$ & $1$&$1\times 10^{-1}$ \\
 && $(r^{13}_{23})^2$ & $1$&$1\times 10^{-4}$ & $1$&$1\times 10^{-1}$ & $9$&$7\times 10^{-2}$ \\\cline{2-9}
& \multirow{3}{*}{$3\sigma$}
 & $(r^{12}_{13})^2$  & $[0$&$49,\, 1.9]$               & $[4$&$2,\, 18]\times 10^{-1}$  & $[0$&$57,\, 2.5]$ \\
 && $(r^{12}_{23})^2$ & $[0$&$47,\, 3.2]\times 10^{-4}$ & $[6$&$9,\, 15]\times 10^{-2}$   & $[0$&$85,\, 1.7] \times 10^{-1}$ \\
 && $(r^{13}_{23})^2$ & $[0$&$48,\, 3.3]\times 10^{-4}$ & $[0$&$83,\, 1.6]\times 10^{-1}$ & $[6$&$8,\, 15] \times 10^{-2}$ \\\hline
\end{tabular}
\caption{ The parameters $r^{ij}_{kl}$ are given for several values of
  neutrino oscillation parameters. SNH and SIH are strict normal and
  strict inverted hierarchy of neutrino masses, respectively. Rows
  labeled as TBM assume the TBM values for $\theta_{12}$ and
  $\theta_{23}$ and the neutrino mass splittings have been fixed to
  their b.f.p.  values taken from \cite{Maltoni:2004ei}. Rows labeled
  as $3\sigma$ take into account current allowed $3\sigma$ ranges of
  neutrino oscillation parameters.  In the first column, $\theta_{13}$
  has been fixed to its TBM value ($s_{13}=0$), while in the second
  and third columns $s_{13}$ has been fixed to its maximum allowed
  value: $(s_{13}^{\rm max})^2=0.050$ at $3\sigma$ C.L. and the Dirac
  phase is fixed to $\delta=0$ and $\delta=\pi$, respectively.}
\label{tab:NHandIH}
\end{center}\end{table}

\begin{table}[htdp]\begin{center}
\begin{tabular}{|c|c|c||r@{.}l|r@{.}l|r@{.}l|}\hline
\multicolumn{3}{|c||}{} & \multicolumn{2}{|c|}{$s_{13}=0$} & \multicolumn{2}{|c|}{$s_{13}=s_{13}^{\rm max},\, \delta=0$} & \multicolumn{2}{|c|}{$s_{13}=s_{13}^{\rm max},\, \delta=\pi$} \\\hline\hline
\multirow{6}{*}{$\begin{array}{c}\textrm{QD}\\\textrm{NH}\end{array}$} 
& \multirow{3}{*}{TBM}
 & $(r^{12}_{13})^2$ & $1$&$0$              & $1$&$3$               & $7$&$7\times 10^{-1}$ \\
&& $(r^{12}_{23})^2$ & $4$&$4\times 10^{-4}$ & $1$&$2\times 10^{-1}$ & $9$&$4\times 10^{-2}$ \\
&& $(r^{13}_{23})^2$ & $4$&$4\times 10^{-4}$ & $9$&$4\times 10^{-2}$ & $1$&$2\times 10^{-1}$ \\\cline{2-9}
& \multirow{3}{*}{$3\sigma$}
 & $(r^{12}_{13})^2$ & $[0$&$49,\, 1.9]$            & $[0$&$63,\, 3.0]$               & $[3$&$5,\, 17]\times 10^{-1}$ \\
&& $(r^{12}_{23})^2$ & $[1$&$8,\, 12]\times 10^{-4}$ & $[0$&$94,\, 1.8]\times 10^{-1}$ & $[6$&$2,\, 15]\times 10^{-2}$ \\
&& $(r^{13}_{23})^2$ & $[1$&$8,\, 12]\times 10^{-4}$ & $[6$&$1,\, 15]\times 10^{-2}$   & $[0$&$93,\, 1.8]\times 10^{-1}$ \\\hline
\multirow{6}{*}{$\begin{array}{c}\textrm{QD}\\\textrm{IH}\end{array}$} 
& \multirow{3}{*}{TBM}
 & $(r^{12}_{13})^2$ & $1$&$0$              & $7$&$6\times 10^{-1}$ & $1$&$3$ \\
&& $(r^{12}_{23})^2$ & $4$&$6\times 10^{-4}$ & $8$&$9\times 10^{-2}$ & $1$&$2\times 10^{-1}$ \\
&& $(r^{13}_{23})^2$ & $4$&$6\times 10^{-4}$ & $1$&$2\times 10^{-1}$ & $8$&$9\times 10^{-2}$ \\\cline{2-9}
& \multirow{3}{*}{$3\sigma$}
 & $(r^{12}_{13})^2$ & $[0$&$49,\, 1.9]$            & $[3$&$4,\, 16]\times 10^{-1}$   & $[0$&$64,\, 3.1]$ \\
&& $(r^{12}_{23})^2$ & $[1$&$9,\, 13]\times 10^{-4}$ & $[5$&$9,\, 15]\times 10^{-2}$   & $[0$&$89,\, 1.8]\times 10^{-1}$ \\
&& $(r^{13}_{23})^2$ & $[1$&$9,\, 13]\times 10^{-4}$ & $[0$&$88,\, 1.7]\times 10^{-1}$ & $[5$&$8,\, 14]\times 10^{-2}$ \\\hline
\end{tabular}
\caption{ The parameters $r^{ij}_{kl}$ are given for several values of
  neutrino oscillation parameters. QD stands for the quasi-degenerate
  limit, while NH (IH) indicate that the neutrino hierarchy is normal
  (inverse). The neutrino parameters have been varied in the same way
  as in Table~\ref{tab:NHandIH}. }
\label{tab:QD}
\end{center}\end{table}

\subsection{Right-handed neutrinos strongly hierarchical}

One can consider the case of degenerate right-handed neutrinos to be
just one extreme limit in a continuum of possibilities. The opposite
extreme case would than be to assume right-handed neutrinos are
strongly hierarchical. Note that here we make the important assumption
that the matrix $R$ is the identity.

\subsubsection{Dominant $M_1$}
\begin{figure}[htb] \centering
\includegraphics[height=5cm,width=.49\linewidth]{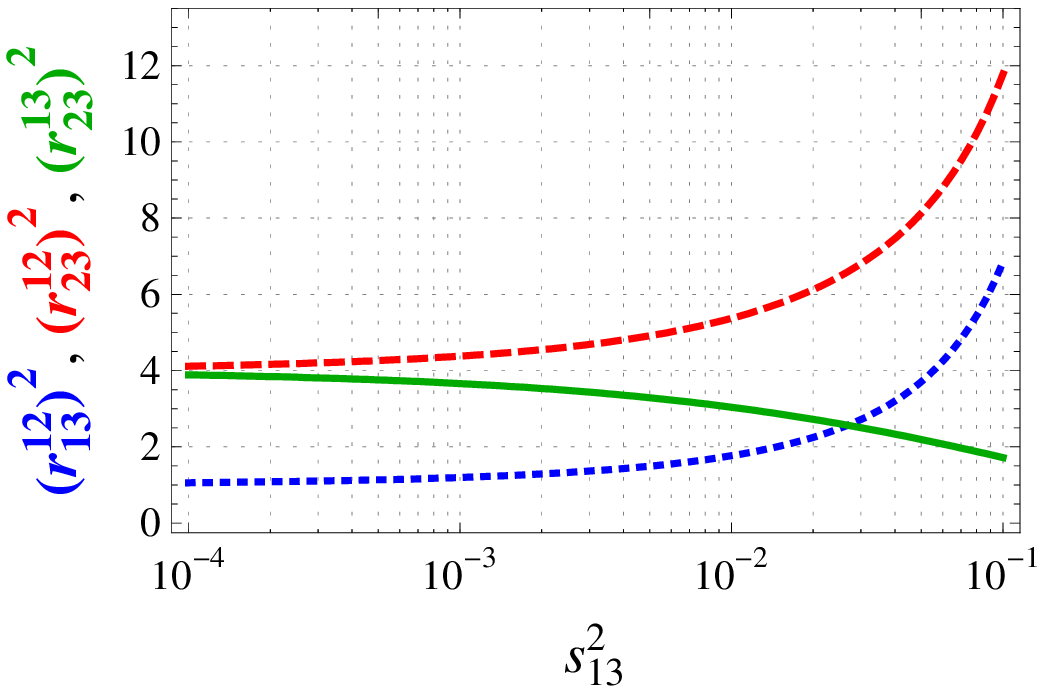}
\includegraphics[height=5cm,width=.49\linewidth]{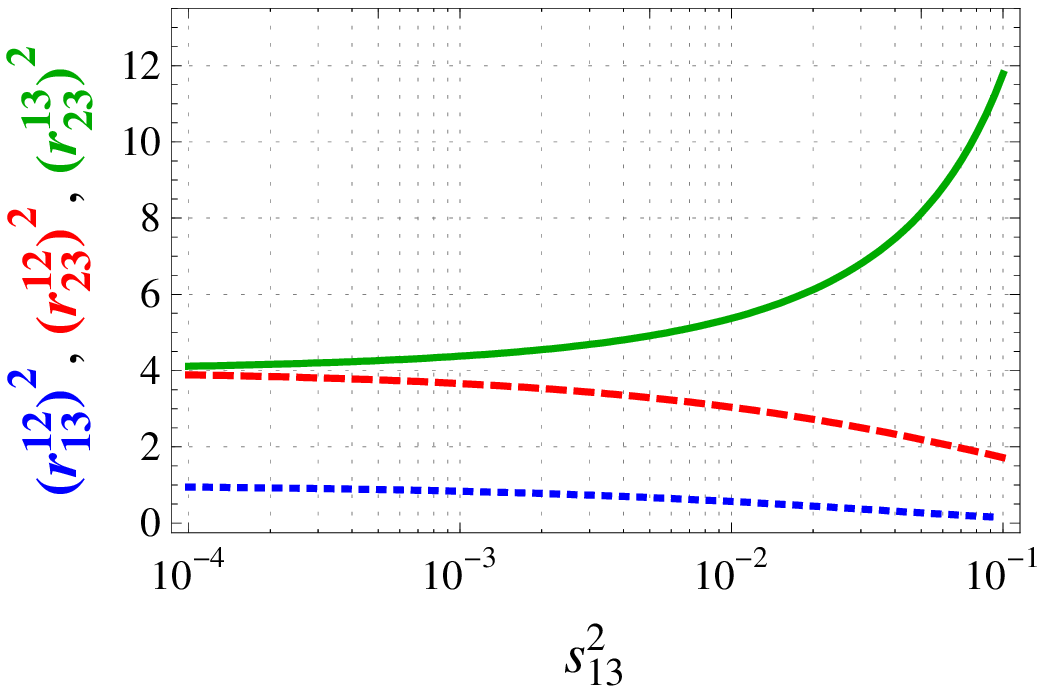}
\caption{Square ratios $(r^{12}_{13})^2$ (blue line, dotted line), 
$(r^{12}_{23})^2$ (red line, dashed line) and $(r^{13}_{23})^2$ (green 
line, full line) versus $s_{13}$ for $\delta=0$ (left panel) and $\delta=\pi$ 
(right panel) for the case of $M_1$ being dominant. The remaining 
neutrino parameters have been fixed to their b.f.p. values.}
\label{fig:m1ana}
\end{figure}

If $M_1$ is the heaviest mass eigenvalue, the leading terms for the
off-diagonal slepton masses are (in case $m_1\ne 0$)
\begin{eqnarray}\label{eq:domM1}
\left(\Delta{M_{\tilde{L}}^2}\right)_{12} & 
\propto & c_{13} c_{12} ( s_{12} c_{23} + s_{13} e^{-i\delta} c_{12} s_{23} ) 
\\ \nonumber
\left(\Delta{M_{\tilde{L}}^2}\right)_{13} & 
\propto & c_{13} c_{12} ( s_{12} s_{23} - s_{13} e^{-i\delta} c_{12} c_{23} ) 
\\ \nonumber
\left(\Delta{M_{\tilde{L}}^2}\right)_{23} & 
\propto & s_{12}^2 s_{23} c_{23} - s_{13} s_{12} c_{12} 
(e^{-i\delta} c_{23}^2 -e^{i\delta} s_{23}^2) - 
s_{13}^2 c_{12}^2 s_{23} c_{23} 
\end{eqnarray}
For the special case of $s_{13}=0$, the ratios simplify to
$r^{12}_{13} = \frac{c_{23}}{s_{23}}$, $r^{12}_{23} =
\frac{c_{12}}{s_{12}s_{23}}$ and $r^{13}_{23}=
\frac{c_{12}}{s_{12}c_{23}}$. Note the large difference in the
numerical values compared to the case of degenerate right-handed
neutrinos. Here, for example for $s_{13}=0$ one finds
$(r^{13}_{23})^2=4$, whereas in the case of degenerate right-handed
neutrinos one obtains $(r^{13}_{23})^2=0.017$ [best fit point (b.f.p.) values for $\Ds$
and $\Da$].  For nonzero values of $s_{13}$ Fig.~\ref{fig:m1ana}
shows that $(r^{ij}_{kl})^2$ depend to a much lesser degree on
$s_{13}$ than for the case of degenerate right-handed
neutrinos. Especially, note that for the case of $M_1$ dominance
considered here none of the $(r^{ij}_{kl})^2$ vanish in the allowed
range of $s_{13}$. Numerical values for extreme values of $s_{13}$ are
summarized in Table~\ref{tab:dominantM}.

\subsubsection{Dominant $M_2$}

\begin{figure}[htb] \centering
\includegraphics[height=5cm,width=.49\linewidth]{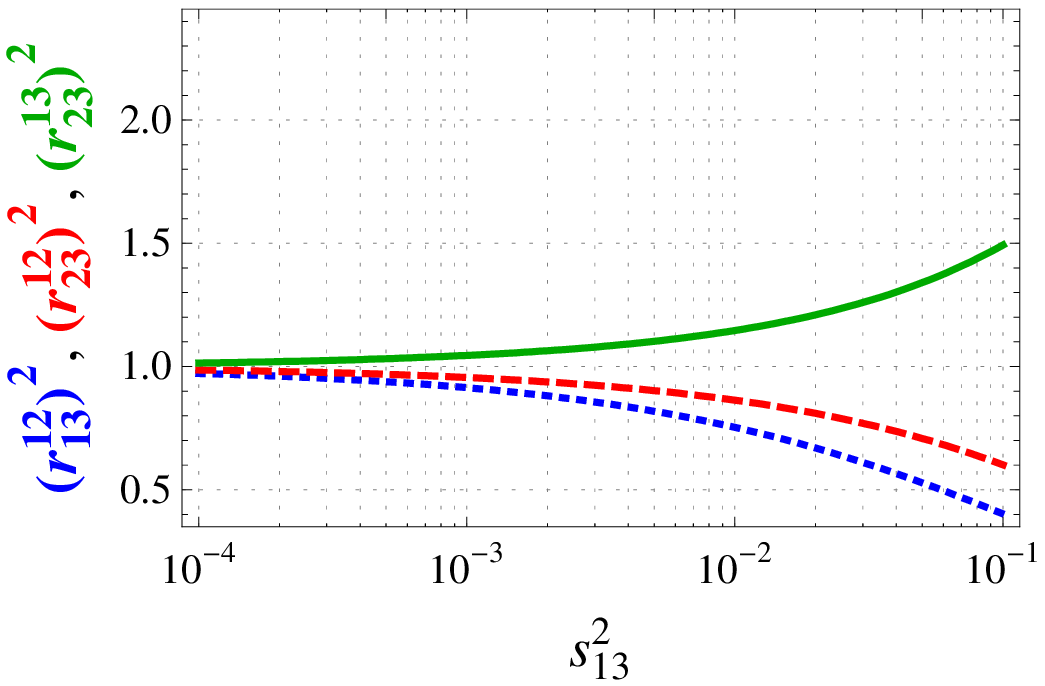}
\includegraphics[height=5cm,width=.49\linewidth]{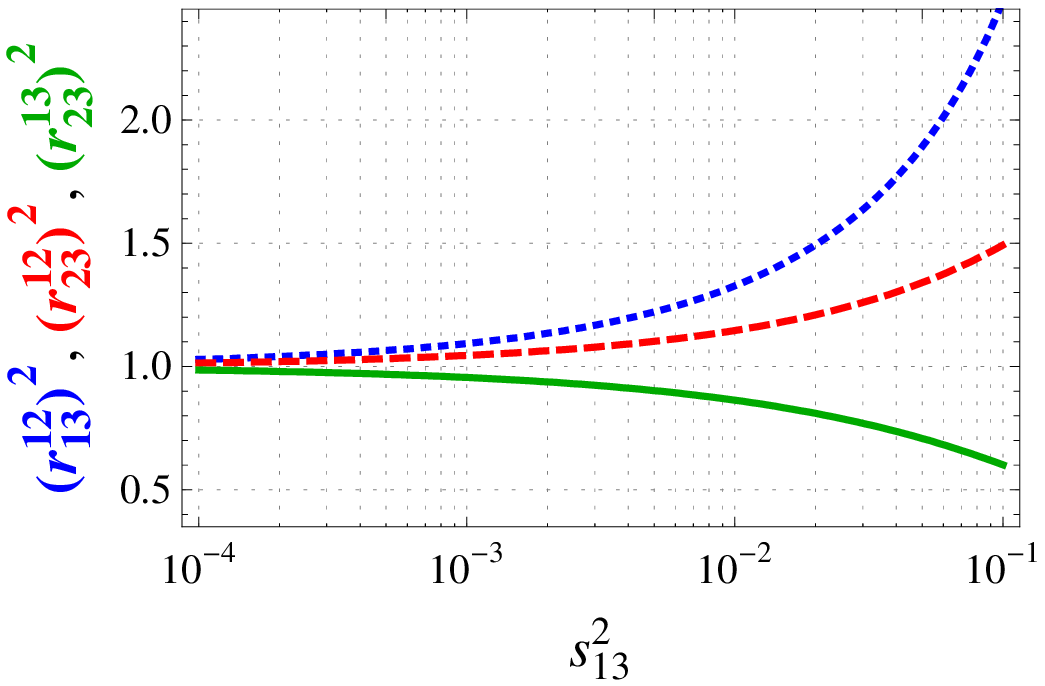}
\caption{Square ratios $(r^{12}_{13})^2$ (blue line, dotted line), 
$(r^{12}_{23})^2$ (red line, dashed line) and $(r^{13}_{23})^2$ (green 
line, full line) versus $s_{13}$ for $\delta=0$ (left panel) and $\delta=\pi$ 
(right panel) in the case where $M_2$ is dominant. The other neutrino 
parameters have been fixed to their b.f.p. values.}
\label{fig:m2ana}
\end{figure}

If $M_2$ is the heaviest mass eigenvalue, the dominant terms for 
the off-diagonal slepton masses are 
\begin{eqnarray}\label{eq:domM2}
  \left(\Delta{M_{\tilde{L}}^2}\right)_{12} & 
  \propto & c_{13} s_{12} ( c_{12} c_{23} - s_{13} e^{-i\delta} s_{12} s_{23} ) 
  \\ \nonumber 
  \left(\Delta{M_{\tilde{L}}^2}\right)_{13} & 
  \propto & c_{13} s_{12} ( c_{12} s_{23} + s_{13} e^{-i\delta} s_{12} c_{23} ) 
  \\ \nonumber 
  \left(\Delta{M_{\tilde{L}}^2}\right)_{23} & 
  \propto & c_{12}^2 s_{23} c_{23} + s_{13} s_{12} c_{12} 
  (e^{-i\delta} c_{23}^2 -e^{i\delta} s_{23}^2) - s_{13}^2 s_{12}^2 s_{23} 
  c_{23}\end{eqnarray}
For the special case of $s_{13}=0$, the ratios simplify to 
$r^{12}_{13} = \frac{c_{23}}{s_{23}}$,  
$r^{12}_{23}= \frac{s_{12}}{c_{12}s_{23}}$ and  
$r^{13}_{23} = \frac{s_{12}}{c_{12}c_{23}}$. Here, for example, for 
$s_{13}=0$ one finds $(r^{13}_{23})^2=1$, whereas for the case of $M_1$ 
being dominant this quantity is expected to be $(r^{13}_{23})^2=4$. 
Figure~\ref{fig:m2ana} shows the $(r^{ij}_{kl})^2$ as function of 
$s_{13}^2$ for the $M_2$ dominance case. Again the dependence on 
$s_{13}$ is weaker than in the case of degenerate right-handed 
neutrinos. As in the previous case $(r^{ij}_{kl})^2$ never 
vanishes in the allowed range of $s_{13}^2$. Finally, the 
numerical values also differ from the ones found for the case of 
$M_1$ dominance. A summary of numerical values for extreme values 
of $s_{13}$ is given in Table~\ref{tab:dominantM}.

\subsubsection{Dominant $M_3$}

If terms proportional to $M_3$ give the leading contribution one finds 
\begin{eqnarray}\label{eq:domM3}
\left(\Delta{M_{\tilde{L}}^2}\right)_{12} & 
\propto & s_{13} e^{-i\delta} c_{13} s_{23} \\ \nonumber
\left(\Delta{M_{\tilde{L}}^2}\right)_{13} & 
\propto & s_{13} e^{-i\delta} c_{13} c_{23} \\ \nonumber
\left(\Delta{M_{\tilde{L}}^2}\right)_{23} & 
\propto & c_{13}^2 s_{23} c_{23} 
\end{eqnarray}
For the special case of $s_{13}=0$, one finds that
$r^{12}_{23}=r^{13}_{23}=0$, otherwise both ratios are proportional to
$s_{13}$. These numerical values allow us to distinguish the $M_3$
dominance case from the previous hierarchical cases already
discussed. Numerical values for extreme values of $s_{13}$ are
summarized in Table~\ref{tab:dominantM}.

\begin{table}[htb] \begin{center}
\begin{tabular}{|c|c|c||r@{.}l|r@{.}l|r@{.}l|}\hline
\multicolumn{3}{|c||}{} & \multicolumn{2}{|c|}{$s_{13}=0$} & \multicolumn{2}{|c|}{$s_{13}=s_{13}^{\rm max},\, \delta=0$} & \multicolumn{2}{|c|}{$s_{13}=s_{13}^{\rm max},\, \delta=\pi$} \\\hline\hline
\multirow{6}{*}{$M_1$} 
& \multirow{3}{*}{TBM}
 & $(r^{12}_{13})^2$  & $1$&$0$ & $3$&$7$ & $2$&$7\times 10^{-1}$ \\
 && $(r^{12}_{23})^2$ & $4$&$0$ & $8$&$1$ & $2$&$2$ \\
 && $(r^{13}_{23})^2$ & $4$&$0$ & $2$&$2$ & $8$&$1$ \\\cline{2-9}
& \multirow{3}{*}{$3\sigma$}
 & $(r^{12}_{13})^2$  & $[0$&$49,\, 1.9]$ & $[1$&$5,\, 14]$ & $[0$&$66,\, 6.6]\times 10^{-1}$ \\
 && $(r^{12}_{23})^2$ & $[2$&$2,\, 8.4]$  & $[3$&$3,\, 35]$ & $[1$&$5,\, 3.4]$ \\
 && $(r^{13}_{23})^2$ & $[2$&$3,\, 8.6]$  & $[1$&$5,\, 35]$ & $[3$&$3,\, 38]$ \\\hline
\multirow{6}{*}{$M_2$} 
& \multirow{3}{*}{TBM}
 & $(r^{12}_{13})^2$  & $1$&$0$ & $5$&$3\times 10^{-1}$ & $1$&$9$ \\
 && $(r^{12}_{23})^2$ & $1$&$0$ & $7$&$1\times 10^{-1}$ & $1$&$3$ \\
 && $(r^{13}_{23})^2$ & $1$&$0$ & $1$&$3$               & $7$&$1\times 10^{-1}$ \\\cline{2-9}
& \multirow{3}{*}{$3\sigma$}
 & $(r^{12}_{13})^2$  & $[0$&$49,\, 1.9]$ & $[2$&$1,\, 11]\times 10^{-1}$  & $[0$&$85,\, 4.5]$ \\
 && $(r^{12}_{23})^2$ & $[0$&$52,\, 2.0]$ & $[4$&$2,\, 12]\times 10^{-1}$   & $[0$&$61,\, 3.4]$ \\
 && $(r^{13}_{23})^2$ & $[0$&$53,\, 2.0]$ & $[0$&$62,\, 3.5]$               & $[4$&$2,\, 12]\times 10^{-1}$ \\\hline
\multicolumn{3}{|c||}{} & \multicolumn{2}{|c|}{$s_{13}=0$} & \multicolumn{2}{|c|}{$s_{13}=s_{13}^{\rm max}$} & \multicolumn{2}{|c}{}\\\cline{1-7}\cline{1-7}
\multirow{6}{*}{$M_3$} 
& \multirow{3}{*}{TBM}
 & $(r^{12}_{13})^2$  & \multicolumn{2}{|l|}{$\ \:-$} & $1$&$0$               & \multicolumn{2}{|c}{} \\
 && $(r^{12}_{23})^2$ & $0$&$0$                       & $1$&$1\times 10^{-1}$ & \multicolumn{2}{|c}{}\\
 && $(r^{13}_{23})^2$ & $0$&$0$                       & $1$&$1\times 10^{-1}$ & \multicolumn{2}{|c}{}\\\cline{2-7}
& \multirow{3}{*}{$3\sigma$}
 & $(r^{12}_{13})^2$  & \multicolumn{2}{|l|}{$\ \:-$} & $[0$&$52,\, 2.0]$               & \multicolumn{2}{|c}{}\\
 && $(r^{12}_{23})^2$ & $0$&$0$                       & $[0$&$80,\, 1.6]\times 10^{-1}$ & \multicolumn{2}{|c}{}\\
 && $(r^{13}_{23})^2$ & $0$&$0$                       & $[0$&$79,\, 1.5]\times 10^{-1}$ & \multicolumn{2}{|c}{}\\\cline{1-7}
\end{tabular}
\caption{ The parameters $r^{ij}_{kl}$ are given for several values of
  neutrino oscillation parameters. Each row labeled as $M_i$ is
  calculated assuming the contribution from neutrino with mass $M_i$
  is dominant. Neutrino oscillation parameters have been varied as in
  Table~\ref{tab:NHandIH}. Notice that the row for dominant $M_3$
  gives the same numerical result for the Dirac phase $\delta=0$ and
  $\delta=\pi$.}
\label{tab:dominantM}
\end{center}\end{table}

\section{Numerical results}
\label{sec:num}

The analytical results presented above allow us to estimate ratios of
branching ratios for LFV decays. For absolute values of the branching
ratios, as well as for cross-checking the reliability of the
analytical estimates, one must resort to a numerical calculation. In
this section we present results of such a numerical calculation. All
results presented below have been obtained with the lepton flavour
violating version of the program package \texttt{SPHENO}~\cite{Porod:2003um}.
For definiteness we will present results only for the mSugra
``standard points'' SPS3 \cite{Allanach:2002nj} and SPS1a'
\cite{AguilarSaavedra:2005pw}, taken as reference examples.  However,
we have checked with a number of other points that our results for
ratios of branching ratios are generally valid.
SPS1a' \cite{AguilarSaavedra:2005pw} is a typical point 
in the ``bulk'' region for SUSY dark matter. It is a slightly modified 
version of the original SPS1a point of \cite{Allanach:2002nj}, which 
gives better agreement with the latest constraints from cold dark matter 
abundance. It has a relatively light slepton spectrum, i.e. left 
sleptons around 200 GeV. SPS3 \cite{Allanach:2002nj} is a point in 
the co-annihilation region for SUSY dark matter. Left sleptons in this point 
are heavier than in SPS1a', i.e. have masses around 350 GeV. We have 
chosen these two points to show the complementarity between low-energy 
searches for LFV and LFV scalar tau decays at the LHC, see 
also the discussion below.

Our numerical procedure to fit the neutrino masses is as follows.
Inverting the seesaw equation, see Eq.~(\ref{meff}), one can get a
first guess of the Yukawa couplings for any fixed values of the light
neutrino masses and mixing angles as a function of the corresponding
right-handed neutrino masses. 
We then run numerically the renormalization group equations taking into account all flavour 
structures in matrix form. We integrate out every right-handed neutrino 
and its superpartner at the scale corresponding to its mass and 
calculated the corresponding contribution to the dimension-5 operator 
which is evaluated to the electroweak scale. This way we  
obtain the exact neutrino masses and mixing angles for this
first guess.  The difference between the results obtained numerically
and the input numbers is then minimized in an iterative procedure
until convergence is achieved. As is well-known neutrino masses and
mixing angles run very little if physical light neutrino masses are
hierarchical \cite{Antusch:2005gp}.  Thus, barring the exceptional
case where neutrinos become very degenerate, one usually reaches
numerical convergence very fast.
For degenerate left neutrinos convergence from first guess to exact
results can be slow, especially for relatively large values for the
right-handed neutrino masses, which require larger Yukawa coupling
constants. In this case we used a numerical fit
procedure~\cite{James:1975dr} based on the program
\texttt{MINUIT}~\footnote{Minimization package from the CERN Program
  Library. Documentation can be found at
  \texttt{http://cernlib.web.cern.ch/cernlib/}}.

In the following two subsections we present numerical results first
for the case of degenerate right-handed neutrinos, then for the
case(s) of very hierarchical right-handed neutrinos. We have checked
numerically that, as expected from Eq.~(\ref{running}), right-sleptons
have small branching ratios for LFV final states. Thus, the discussion
concentrates on the decays of the ``left'' staus ${\tilde\tau}_2\simeq
{\tilde\tau}_L$.

\subsection{Degenerate right-handed neutrinos}

\begin{figure}[htbp]
\begin{center}
\vspace{5mm}
\includegraphics[width=80mm,height=60mm]{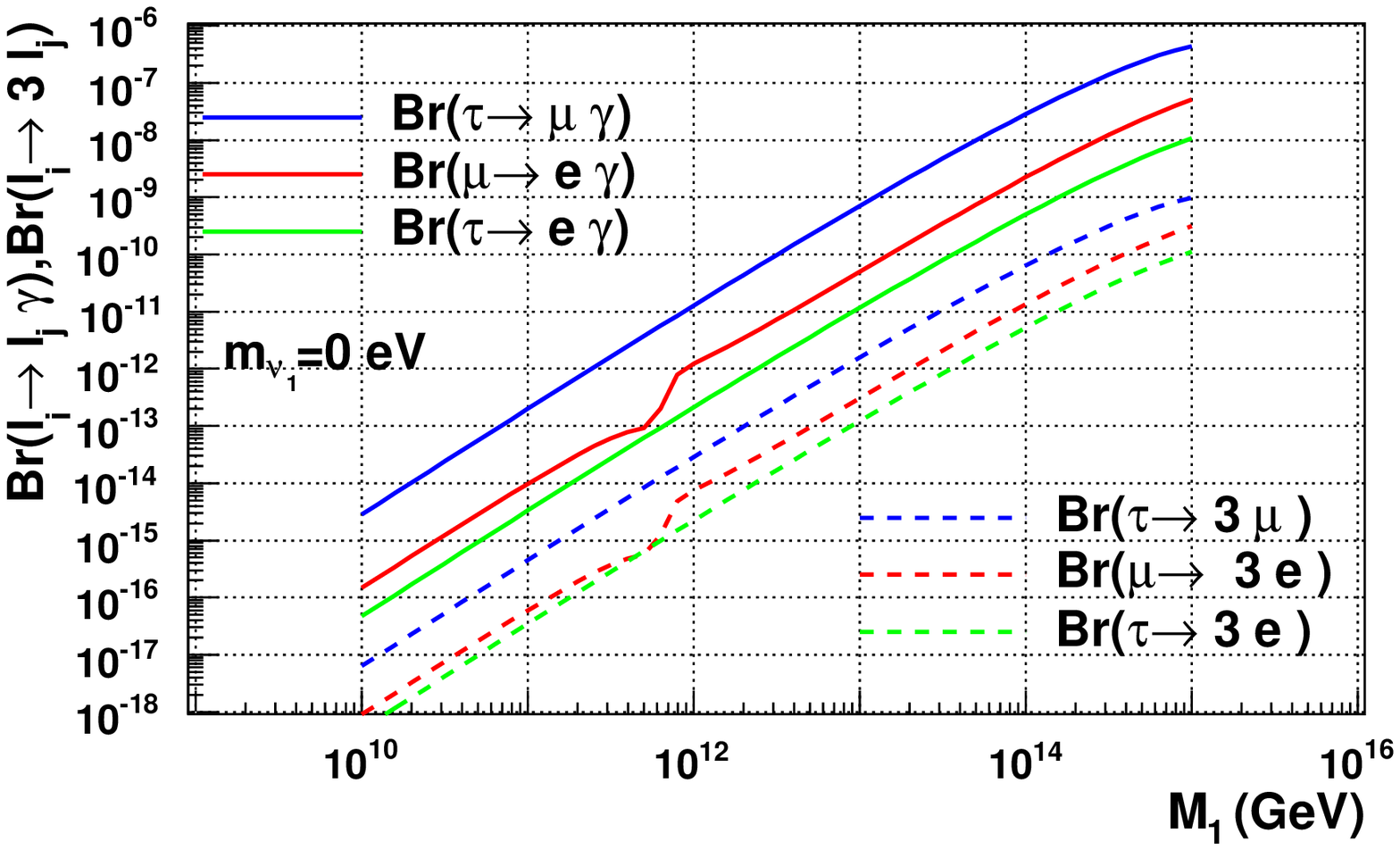}
\includegraphics[width=80mm,height=60mm]{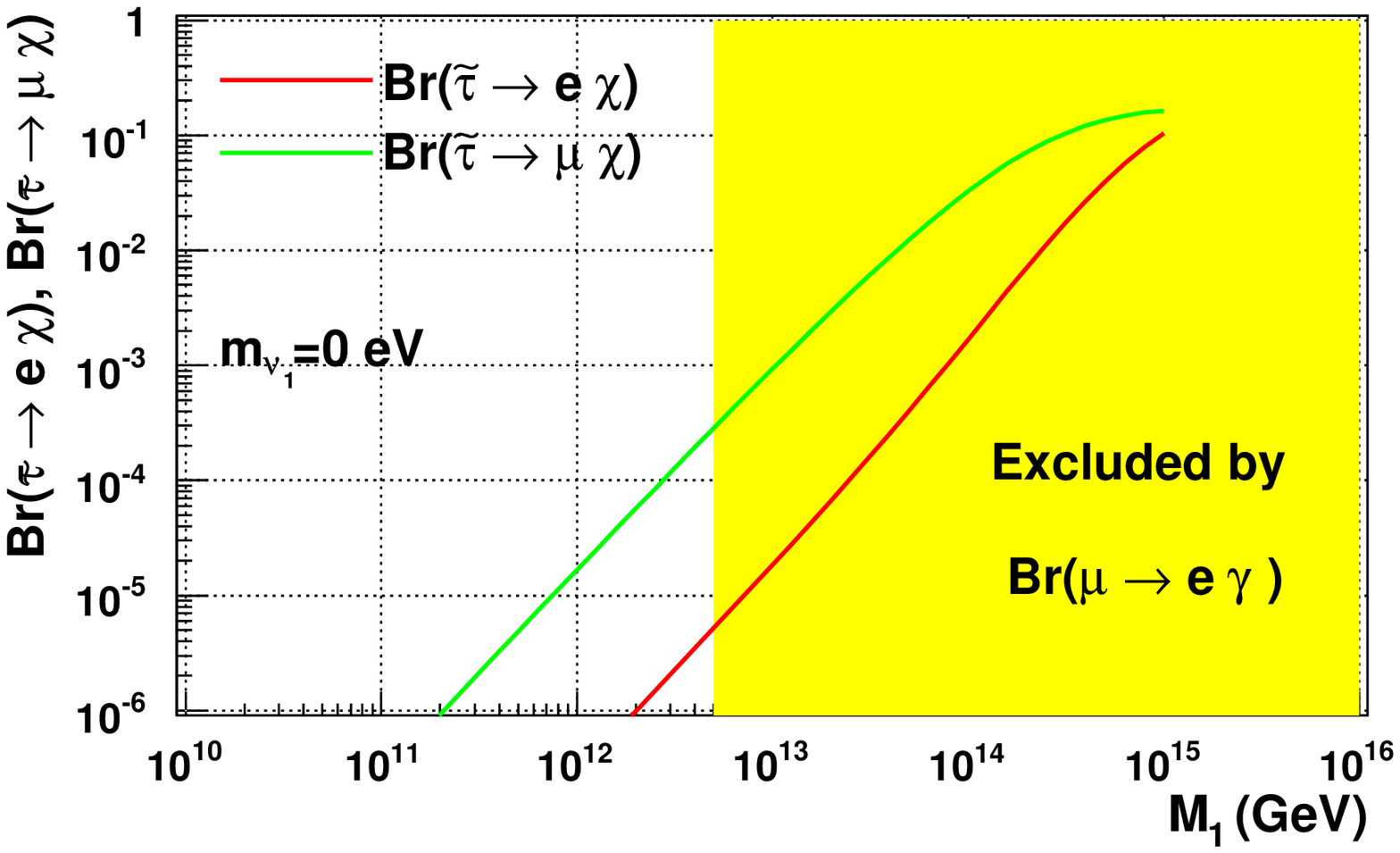}
\end{center}
\caption{Branching ratios for $l_i \to l_j + \gamma$ and $l_i \to 3
  l_j$ (left) and ${\tilde\tau}_2 \to e +\chi^0_1$ and ${\tilde\tau}_2 
  \to \mu +\chi^0_1$ (right) for the standard point SPS1a' versus $M_R$,
  assuming degenerate right-handed neutrinos. Neutrino oscillation
  parameters have been fixed to the best fit values for $\Ds$ and
  $\Da$, with exact tribimaximal neutrino angles. We also set
  $m_1=0$.  The coloured region in the right-side plot is excluded
  from the current experimental limit on Br($\mu \to e +
  \gamma$). Thus, one expects for SPS1a' only very small branching
  ratios for LFV scalar tau decays. (Compare to Fig.~(\ref{fig:BrSPS3}).}
\label{fig:BrSPS1ap}
\end{figure}

\begin{figure}[htbp]
\begin{center}
\vspace{5mm}
\includegraphics[width=80mm,height=60mm]{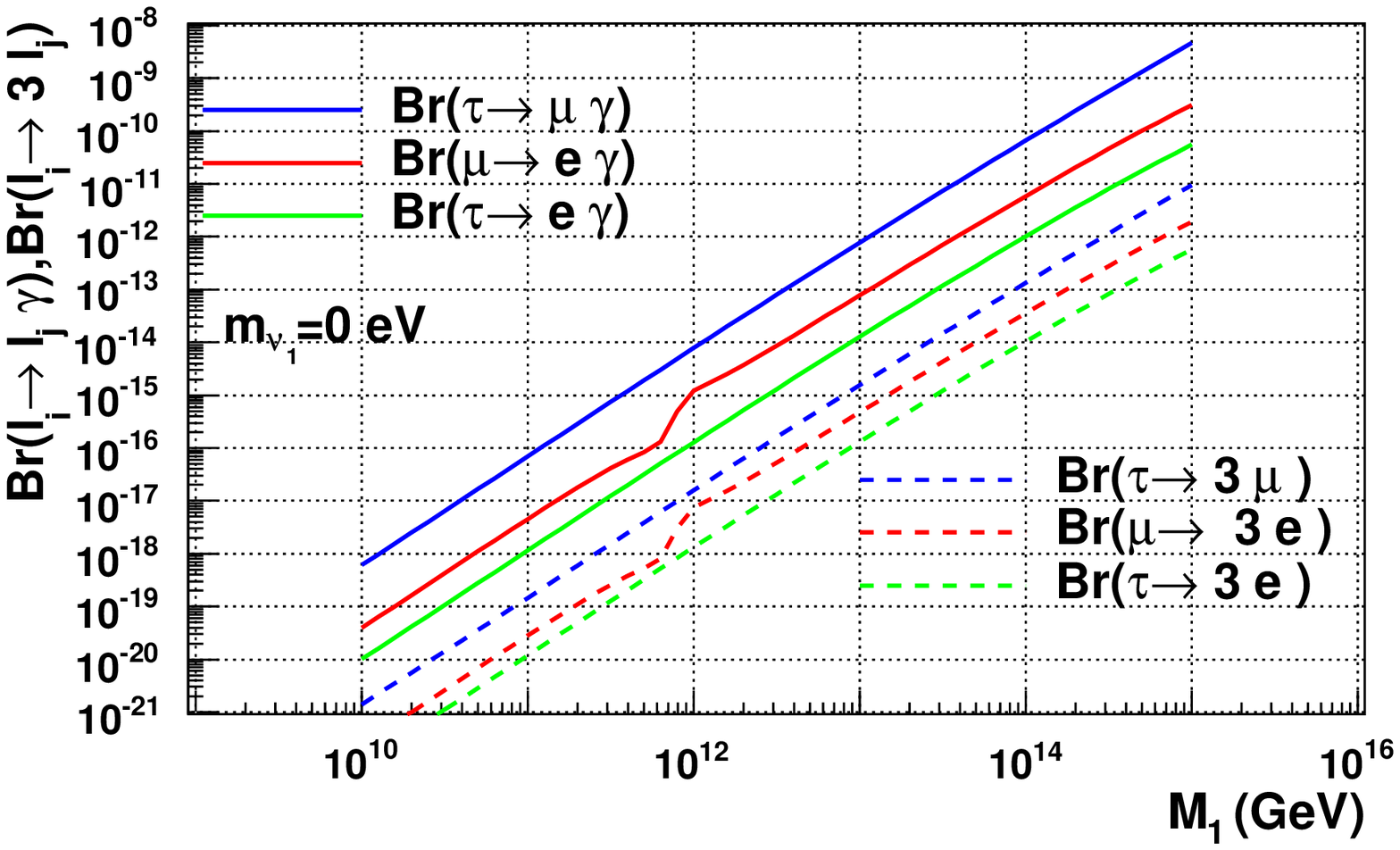}
\includegraphics[width=80mm,height=60mm]{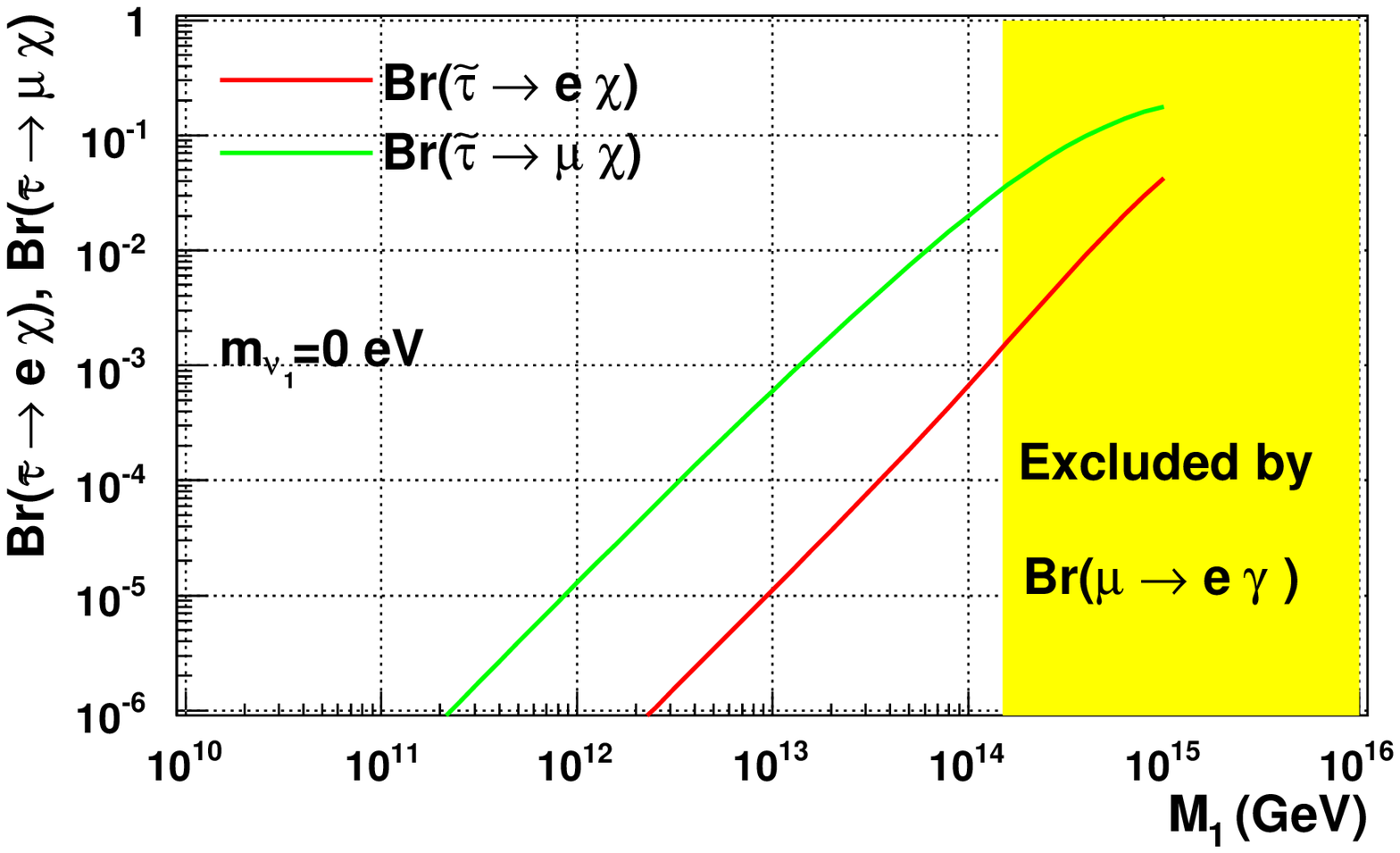}
\end{center}
\caption{Same as Fig.~\ref{fig:BrSPS1ap}, but for the mSugra
  standard point SPS3. In this point the constraints on the LFV
  ${\tilde\tau}_2$ decays from the upper limit on $\mu \to e + \gamma$
  are much less severe than for SPS1a'. As a result Br(${\tilde\tau}_2
  \to \mu + \chi^0_1$) could be as large as several percent with all
  low-energy constraints fulfilled.}
\label{fig:BrSPS3}
\end{figure}

In this subsection we still adopt the simplifying ansatz that $R=1$,
see Eq.~(\ref{Ynu}). Two examples for hierarchical light neutrinos are
shown in Fig.~\ref{fig:BrSPS1ap} and Fig.~\ref{fig:BrSPS3}.
Figure~\ref{fig:BrSPS1ap} has the mSugra parameters fixed to the
standard values SPS1a' \cite{Allanach:2002nj,AguilarSaavedra:2005pw},
while Fig.~\ref{fig:BrSPS3} corresponds to SPS3
\cite{Allanach:2002nj}.  The neutrino oscillation data are fitted for
the strict normal hierarchy (SNH) case where $m_1\equiv 0$ with exact
tribimaximal mixing. The plot on the left panel shows low-energy
lepton flavour violating decay branching ratios for $l_i \to
l_j+\gamma$ and $l_i \to 3 l_j$, while the one on the right panel
gives LFV stau (${\tilde\tau}_2$) decay branching ratios as a function
of the right-handed neutrino mass scale $M_1=M_R$.

As expected, all LFV processes show a strong dependence on $M_R$. This
can be straightforwardly understood from Eqs. (\ref{meff}) and
(\ref{running}). Keeping the light neutrino masses constant
$\Delta{M_{\tilde{L}}^2}$ are proportional to $M_R\log M_R$, thus all
LFV branching ratios grow as $(M_R\log M_R)^2$.  As the figures show,
as long as $M_R$ is not too large, all lepton flavour violating
processes show the same dependence on $M_R$.  Ratios of branching
ratios follow very nicely the corresponding analytically calculated
ratios for $(r^{ij}_{kl})^2$, once the corresponding correction
factors are taken into account for the low-energy observables. As is
well
known~\cite{Hisano:1998fj,Arganda:2005ji,Deppisch:2005zm}, 
for most parts of the mSugra parameter space one expects
\begin{equation}\label{relateLFV}
\frac{{\rm Br}(l_i \to 3 l_j)}{{\rm Br}(l_i \to l_j + \gamma)} \simeq 
\frac{\alpha}{3\pi}\Big(\log(\frac{m_{l_i}^2}{m_{l_j}^2})-\frac{11}{4}\Big).
\end{equation}
thus the photonic penguin diagram dominates the three-lepton
decay modes $l_i \to 3 l_j$.

Figures~\ref{fig:BrSPS1ap} and \ref{fig:BrSPS3} do indeed confirm the
validity of this approximation.  Only at large values of $M_R$ one
observes some deviations from the analytical estimates. The reason for
this departure is that in this parameter range the small-angle
approximation no longer holds, as can be seen from the absolute values
for the decay Br(${\tilde\tau}_2 \to \mu + \chi^0_1$), which can reach
more than 10 \% for $M_R\ge 10^{14}$ GeV.
However, Figs.~\ref{fig:BrSPS1ap} and \ref{fig:BrSPS3} also show
how the LFV ${\tilde\tau}_2$ decays are strongly constrained by low
energy data. For the degenerate right-handed neutrino case shown here
(and for $s_{13}=0$), independent of the mSugra parameters Br($\mu \to
e+\gamma$) is the most important constraint. Applying the current
experimental limit on Br($\mu \to e+\gamma$) of Br($\mu \to e +
\gamma$) $\le 1.2 \times 10^{-11}$ \cite{Yao:2006px}, the branching
ratio for Br(${\tilde\tau}_2 \to \mu +\chi^0_1$) is expected to lie
below $10^{-3}$ for SPS1a', whereas it can reach several percent in
case of SPS3. Note that in the range of $M_R$ not excluded by the
limit on Br($\mu \to e+\gamma$) the ratio Br(${\tilde\tau}_2 \to e
+\chi^0_1$)/Br(${\tilde\tau}_2 \to \mu +\chi^0_1$) follows very well
the analytical estimate of Eq.~(\ref{tbmSNH}). The huge difference in
the upper limit for Br(${\tilde\tau}_2 \to \mu +\chi^0_1$) when going
from SPS1a' to SPS3 can be understood from the fact that both
left-sleptons as well as (lightest) neutralino and chargino are
approximately a factor of two heavier for SPS3 than for SPS1a'.  Since
Br$(\mu \to e+\gamma)\propto 1/m_{SUSY}^8$ \cite{Hisano:1995cp} one
expects Br$(\mu \to e+\gamma)$ to be a factor of more than several
hundred lower for SPS3 than for SPS1a'.

\begin{figure}
\begin{center}
\includegraphics[width=80mm, height=60mm]{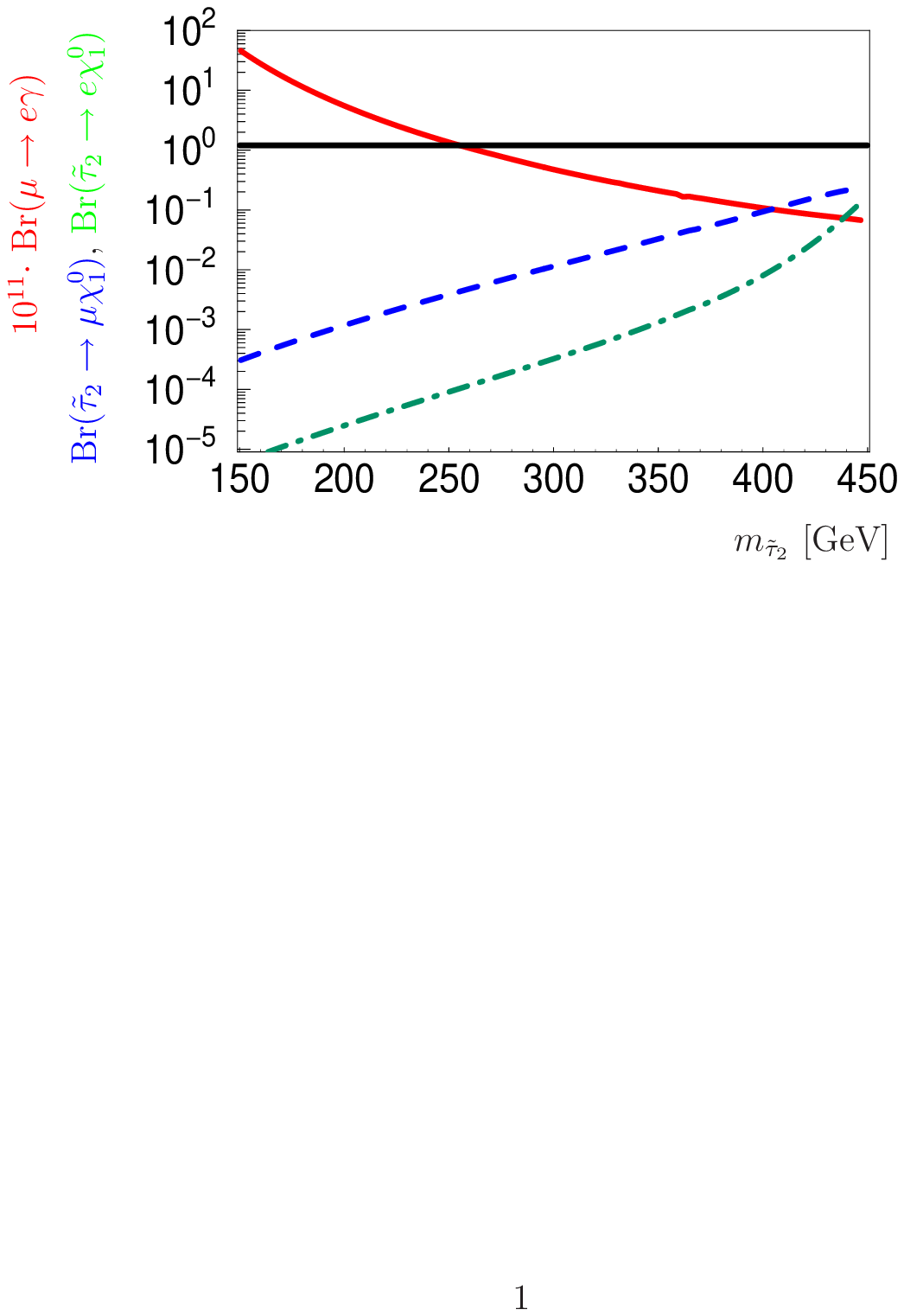}
\end{center}
\vskip1mm
\caption{Branching ratios as function of scalar tau mass. The full
  line (red) is $10^{11} \cdot$ Br($\mu \to  e+\gamma$), the
  dashed line (blue) Br(${\tilde\tau_2} \to  \mu+\chi^0_1$) and
  the dot-dashed line (green) is Br(${\tilde\tau_2} \to 
  e+\chi^0_1$).  Data calculated for SPS1a with parameters varied
  along the ``slope''. Note that SPS1a is used in this 
  plot instead of 
  SPS1a', since for SPSa1' no slope is given in \cite{AguilarSaavedra:2005pw}.
  Right-handed neutrino mass is fixed to $M_R =
  3\times 10^{13}$ GeV.  The black line is the current upper limit on
  Br($\mu \to  e\gamma$).  While SPS1a with $M_R = 3\times
  10^{13}$ GeV is excluded by Br($\mu \to e\gamma$), for slightly
  heavier slepton masses the low-energy constrained can be evaded,
  having at the same time sizeable lepton flavour violating slepton
  decay branching ratios.}
\label{fig:Brs}
\end{figure}

The strong dependence of Br($\mu \to e+ \gamma$) on the supersymmetric
mass spectrum is also seen in Fig.~\ref{fig:Brs}, where we plot
Br($\mu \to e+\gamma$), Br(${\tilde\tau_2} \to \mu+ \chi^0_1$) and
Br(${\tilde\tau_2} \to e +\chi^0_1$) versus the mass of
${\tilde\tau}_2$, for light neutrino parameters as before and a fixed
value of $M_R = 3\times 10^{13}$ GeV. Here, the parameters for the
point SPS1a have been varied around the slope given in
Ref.~\cite{Allanach:2002nj}. Note that Br($\mu \to e+ \gamma$) drops
below the current experimental limit for $m_{{\tilde\tau}_2}$ larger
than about $250$ GeV.
In contrast, the ${\tilde\tau}_2$ LFV decay branching ratios increase
for increasing $m_{{\tilde\tau}_2}$. This is due to the fact that
left-sleptons become more degenerate when $m_0$ is increased along the
slope for SPS1a.
The more degenerate sleptons are, the larger the resulting LFV
parameters, for given light neutrino parameters.
Note, however, that the ratio Br(${\tilde\tau_2} \to e + \chi^0_1$)/
Br(${\tilde\tau_2} \to \mu +\chi^0_1$) remains constant in agreement
with the analytical estimate, as long as Br(${\tilde\tau_2} \to \mu
+\chi^0_1$) is smaller than a few percent. Again this reflects the
fact that the small-angle approximation is valid only for small
branching ratios in the LFV decays.

\begin{figure}[htbp]
\begin{center}
\vspace{5mm}
\includegraphics[width=80mm,height=70mm]{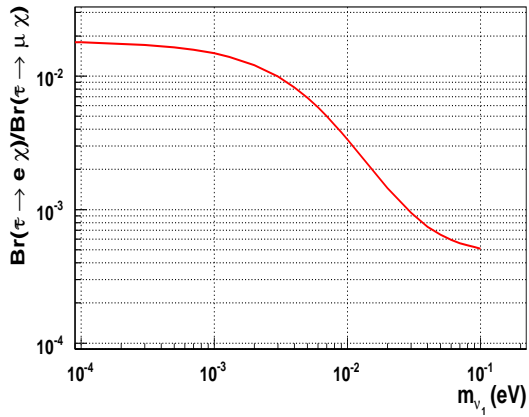}
\end{center}
\caption{Numerically calculated Br(${\tilde\tau}_2 \to e
  +\chi^0_1$)/Br(${\tilde\tau}_2 \to \mu +\chi^0_1$) for the standard point
  SPS1a' versus lightest neutrino mass for the case of normal hierarchy. 
  (compare to Fig.~\ref{fig:degana}).}
\label{fig:vmnu1}
\end{figure}

We have also checked numerically the reliability of our analytical
calculation for the case of $m_1\ne 0$. An example is shown in
Fig.~\ref{fig:vmnu1}. Here we have fixed the mSugra parameters to
the standard point SPS1a', the right-handed neutrino mass scale to
$M_R = 5\times 10^{12}$ GeV, the light neutrino mixing angles to the
TBM values, $\Da$ and $\Ds$ to their b.f.p. values and we have
calculated Br(${\tilde\tau}_2 \to e +\chi^0_1$)/Br(${\tilde\tau}_2 
\to \mu +\chi^0_1$) as a function of the lightest neutrino mass. As shown 
in Fig.~\ref{fig:vmnu1} the value of this ratio obtained within a full
numerical calculation follows very closely the central value given in
Fig.~\ref{fig:degana}, as expected (here we assumed the case of
normal hierarchy).

\subsection{Hierarchical right-handed neutrinos}

Now we turn to the extreme case of very hierarchical right-handed
neutrinos. Again our goal is to check the reliability of the
analytical calculation for this case.
In all figures presented in this subsection we have taken two of the
three right-handed neutrino masses to be constant at $M_R=10^{10}$ GeV
and varied the remaining third right-handed neutrino mass in the
ranges given in the figures. In all cases we have fixed the neutrino
angles to the TBM values, $\Da$ and $\Ds$ to their best-fit values and
assumed normal hierarchical neutrinos. The remaining free parameter
$m_1$ is given in each figure.

\begin{figure}[htbp]
\begin{center}
\includegraphics[width=80mm,height=60mm]{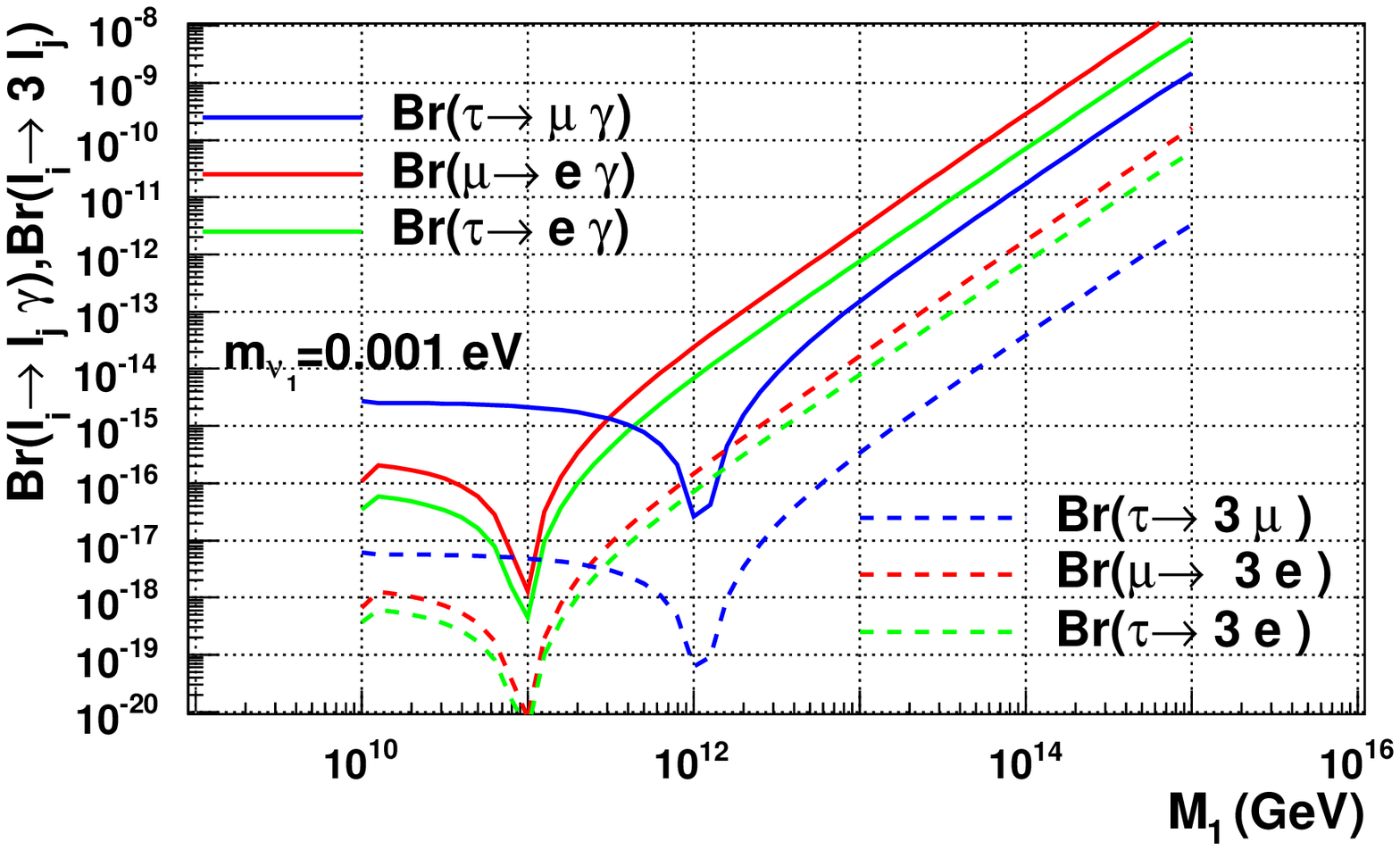}
\includegraphics[width=80mm,height=60mm]{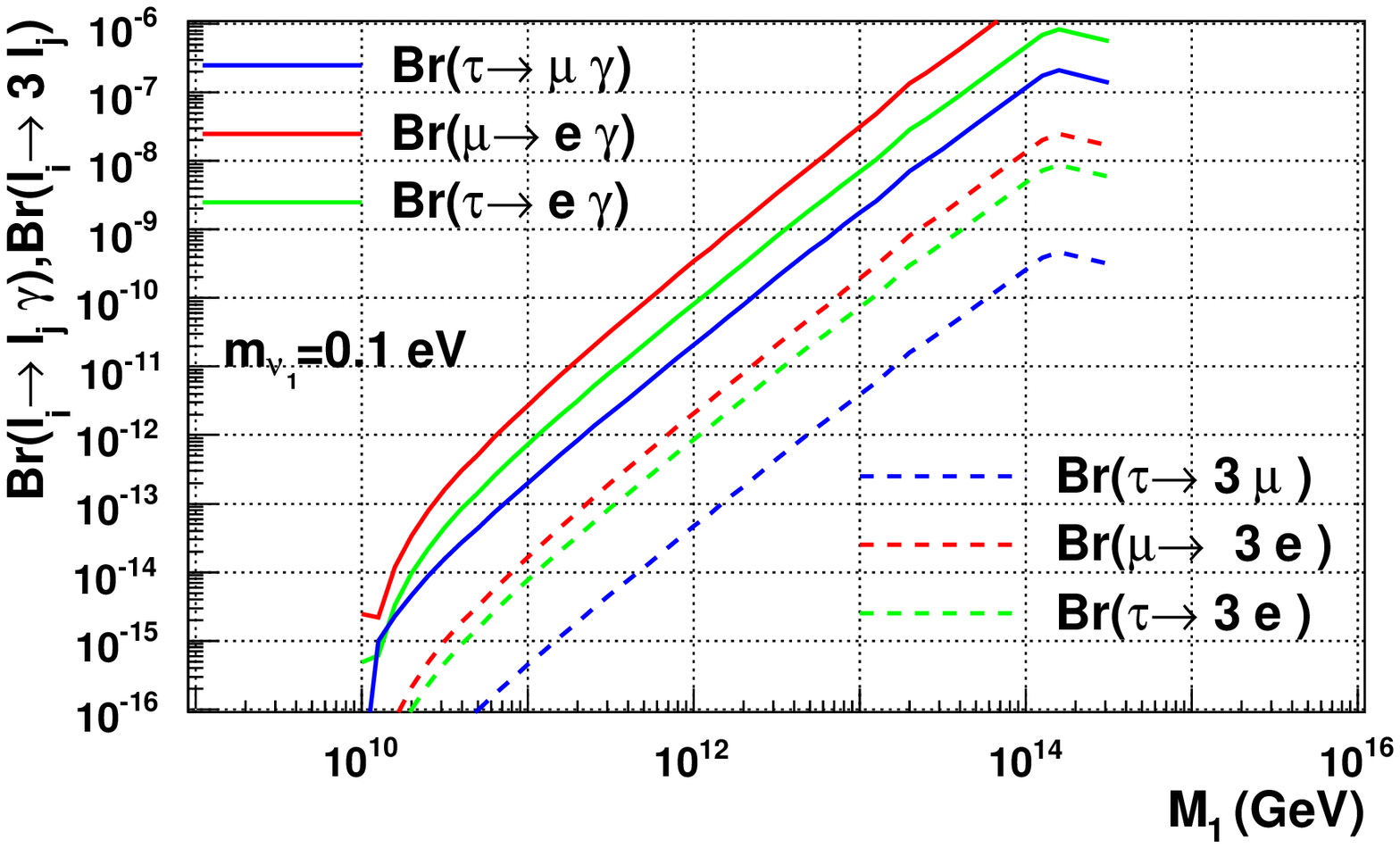}
\end{center}
\vskip1mm
\caption{Branching ratios for $l_i\to  l_j + \gamma $ and 
$l_i \to  3 l_j$, as a function of $M_1$ for constant 
$M_2=M_3=10^{10}$ GeV  and $m_{1}=0.001$ 
eV (left) and for $m_{1}=0.1$ eV (right). mSugra parameters have been 
fixed to SPS1a'.}
\label{fig:AllLFV_M1}
\end{figure}

Figure~\ref{fig:AllLFV_M1} shows LFV lepton decays as a function of
$M_1$ for $m_{1}=0.001$ eV (left) and for $m_{1}=0.1$ eV (right) for
the mSugra parameters fixed at SPS1a'. For $m_{1}=0.001$ eV, the
curves are not monotonous functions of $M_1$. In fact, in the left
figure only for $M_1 \gsim 10^{12}$ GeV do the different branching
fractions follow the analytical estimates of
Eq.~(\ref{eq:domM1}). This is due to the fact that the different
contributions of the $M_i$ to $\Delta{M_{\tilde{L}}^2}_{ij}$ scale
like $m_i M_i\log M_i$, i.e. $M_1$ becomes dominant in the expressions
for the $\Delta{M_{\tilde{L}}^2}_{ij}$ only if $M_1/M_j \gg
m_j/m_1$. This is confirmed by the figure in the right panel, for
which $m_{1}=0.1$ eV has been chosen. Here, the contribution from
$M_1$ to the $\Delta{M_{\tilde{L}}^2}_{ij}$ is indeed the dominant one
for $M_1 \ge$ (few) $\times 10^{10}$ GeV.

\begin{figure}[htbp]
\begin{center}
\includegraphics[width=80mm,height=60mm]{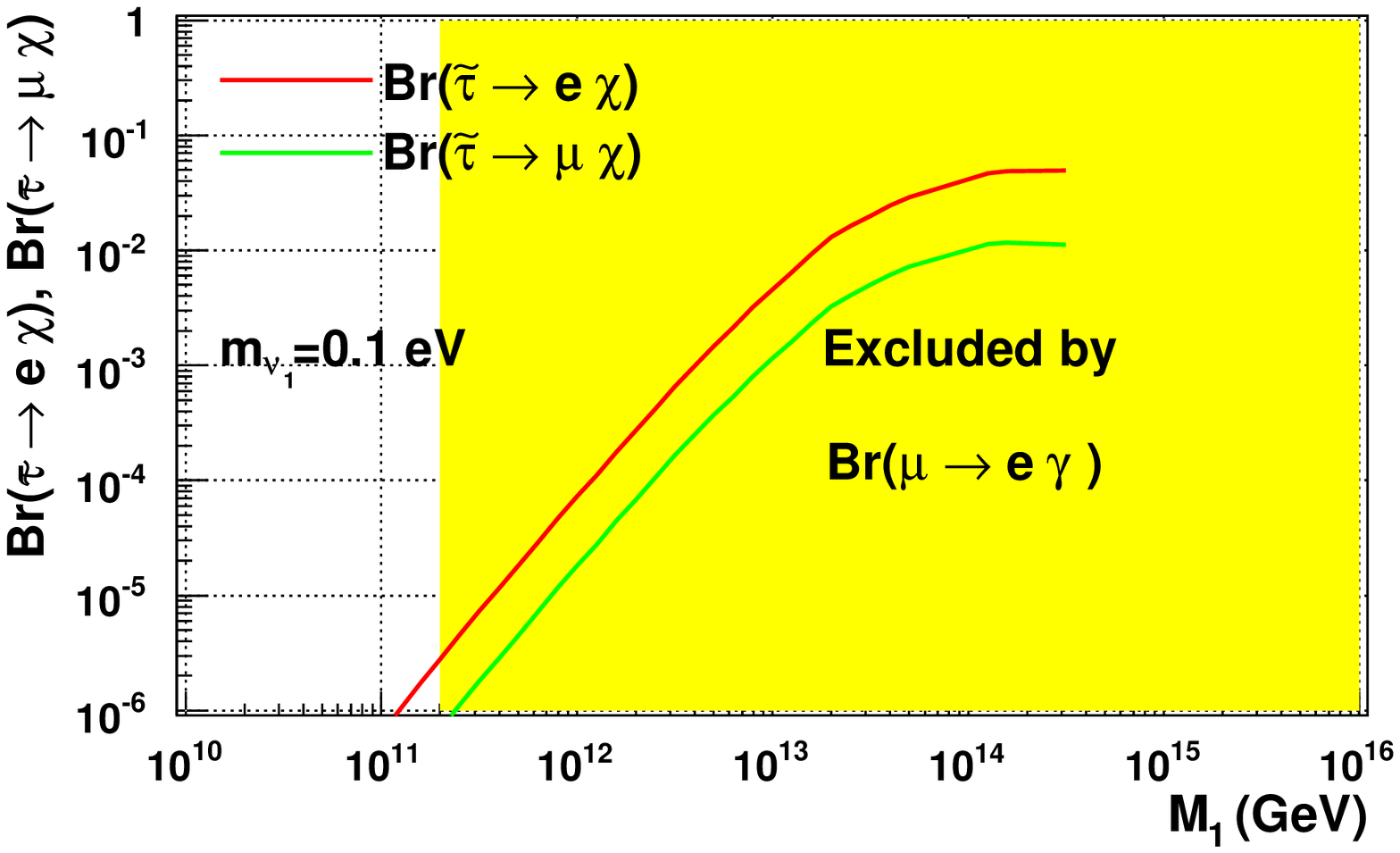}
\includegraphics[width=80mm,height=60mm]{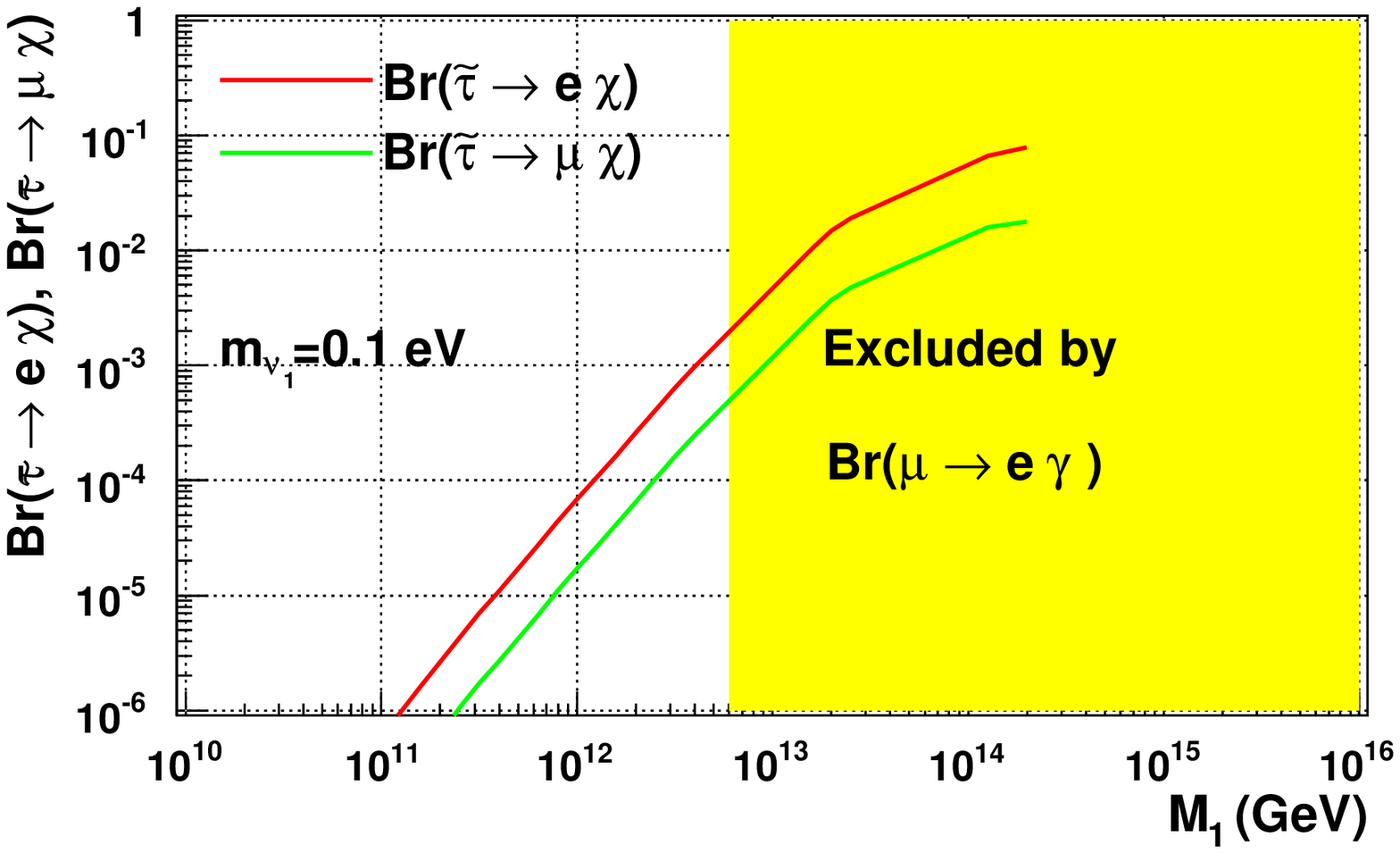}
\end{center}
\vspace{-5mm}
\caption{Branching ratios for $\tilde{\tau}_2 \to   e (\mu) +\chi_1^0$ 
as a function of $M_1$ for constant $M_2=M_3=10^{10}$ GeV for  SPS1a' 
(left) and SPS3 (right).}
\label{fig:StausLFV_M1}
\end{figure}

Figure~\ref{fig:StausLFV_M1} shows branching ratios for $\tilde{\tau}_2 
\to   e (\mu) +\chi_1^0$ as a function of $M_1$ for the two 
mSugra points SPS1a' (left) and SPS3 (right). Again the region excluded 
by the current upper limit on Br$(\mu \to e+\gamma)$ is indicated. 
Ratios of the LVF slepton decays follow the analytical estimate very 
well everywhere in the region allowed by the upper limit on 
Br$(\mu \to e+\gamma)$. One observes, as is the case also for degenerate 
right-handed neutrinos, that for SPS1a' the absolute values for the 
LFV branching ratios are too small to be observable, whereas for the 
mSugra point SPS3 much larger values for LFV scalar tau decays are 
allowed. Note that 
Br$(\tilde{\tau}_2 \to   e +\chi_1^0)$ is larger than 
Br$(\tilde{\tau}_2 \to   \mu +\chi_1^0)$ for $M_1$ dominance, 
in contrast with the case of degenerate right-handed neutrinos.

\begin{figure}[htbp]
\begin{center}
\includegraphics[width=80mm,height=60mm]{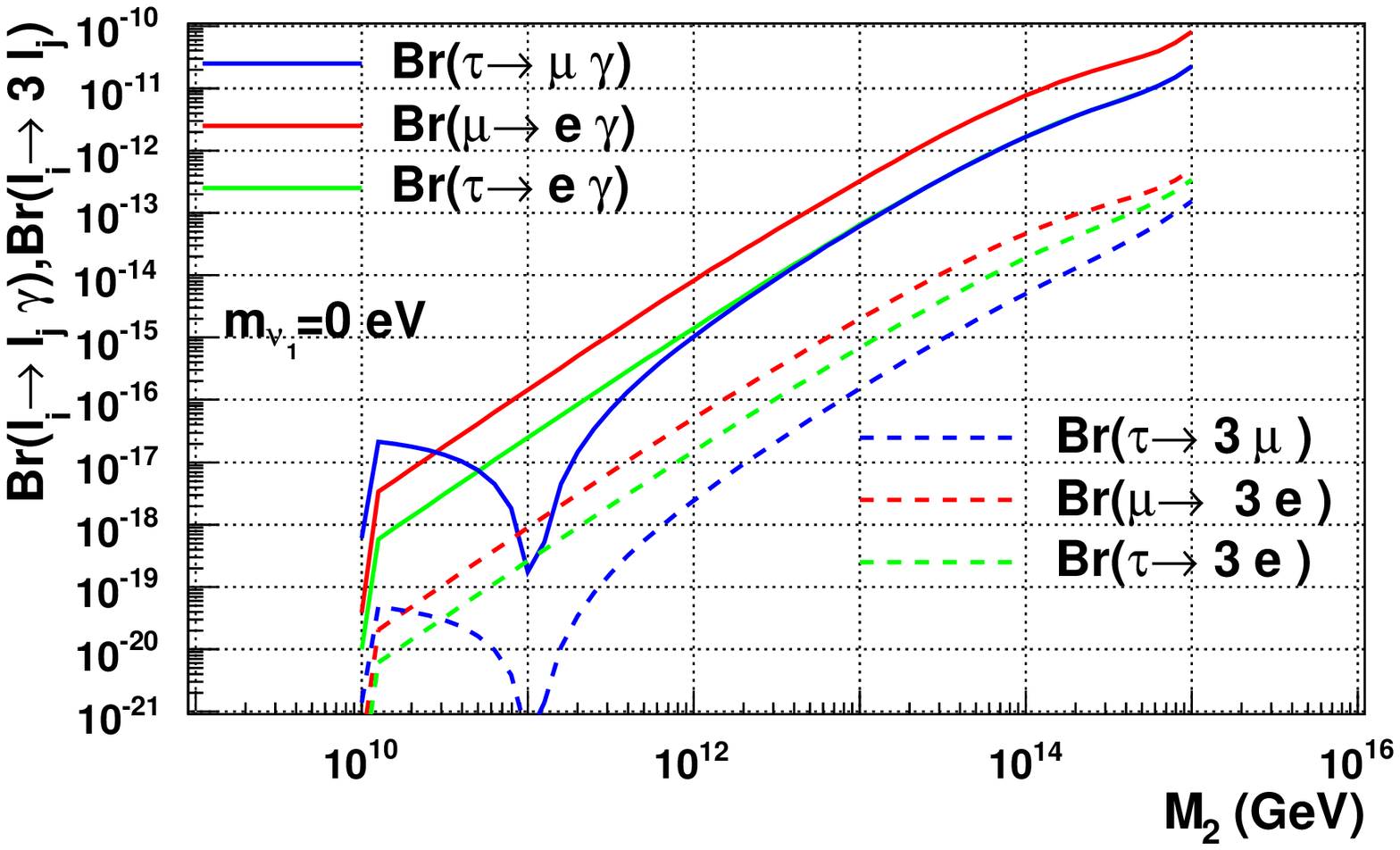}
\includegraphics[width=80mm,height=60mm]{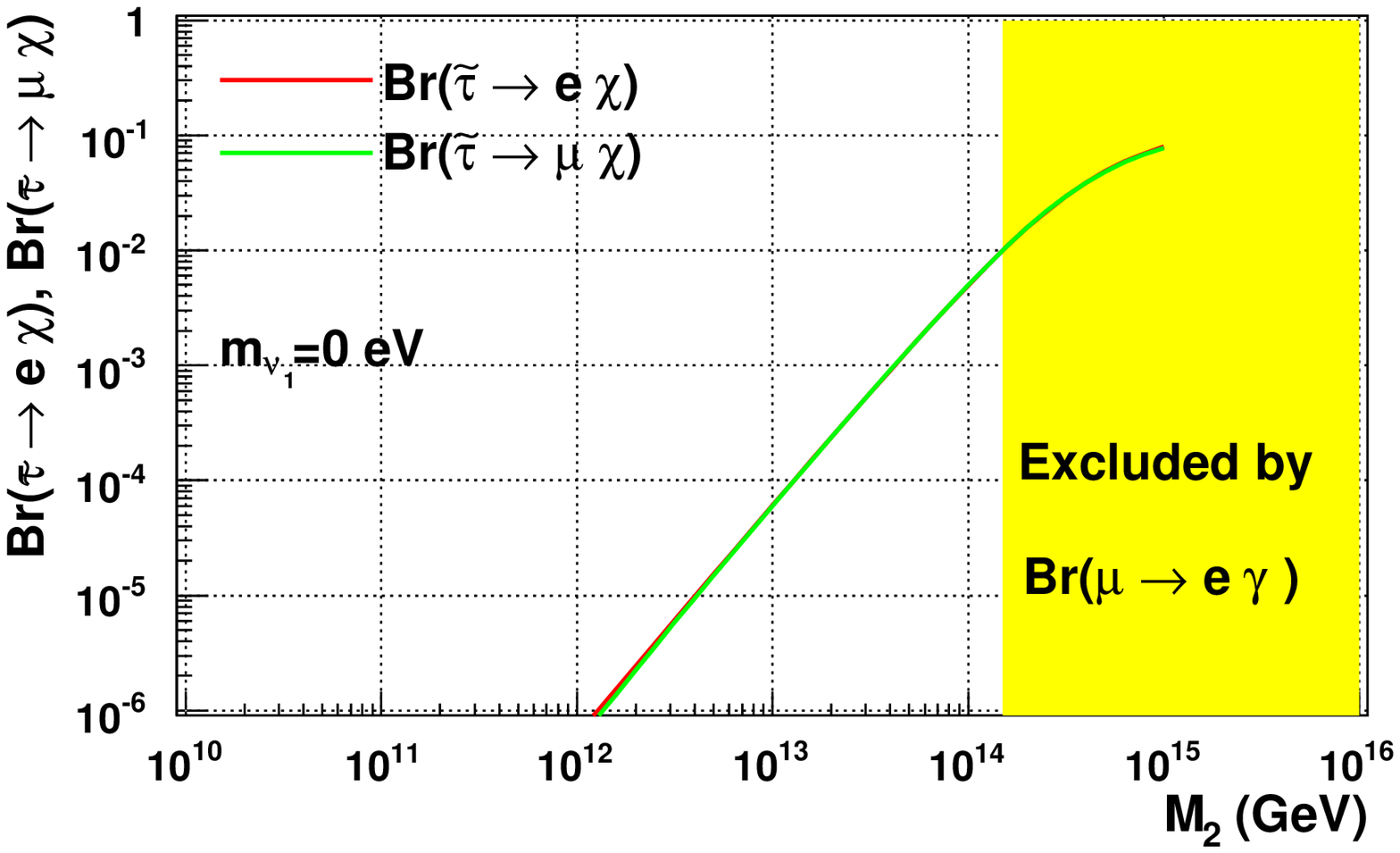}
\end{center}
\vspace{-5mm}
\caption{Branching ratios for $l_i \to  l_j + \gamma $  and 
$l_i \to 3 l_j$ (left) and LFV stau decays (right), for the standard 
point SPS3 as a function of $M_2$ for constant $M_1=M_3=10^{10}$ GeV.} 
\label{fig:LVF_M2}
\end{figure}

Figure~\ref{fig:LVF_M2} shows branching ratios for $l_i \to l_j +
\gamma $ and $l_i \to 3 l_j$ (left) and LFV stau decays (right), for
the standard point SPS3 as a function of $M_2$. As in
Fig.~\ref{fig:AllLFV_M1}, the left panel illustrates that only for
$M_2 \gsim 10^{12}$ GeV the contribution from $M_2$ to the LFV mixing
angles is dominant. For $M_2 \gsim 10^{12}$ GeV the ratios of
branching ratios follow the expectation of Eq.~(\ref{eq:domM2}).  LFV
scalar tau decays as large as 1 \% are allowed in this example.  Note
also that Br$(\tilde{\tau}_2 \to e +\chi_1^0)=$ Br$(\tilde{\tau}_2 \to
\mu +\chi_1^0)$ for $M_2$ dominance and TBM neutrino angles.

\begin{figure}[htbp]
\begin{center}
\includegraphics[width=80mm,height=60mm]{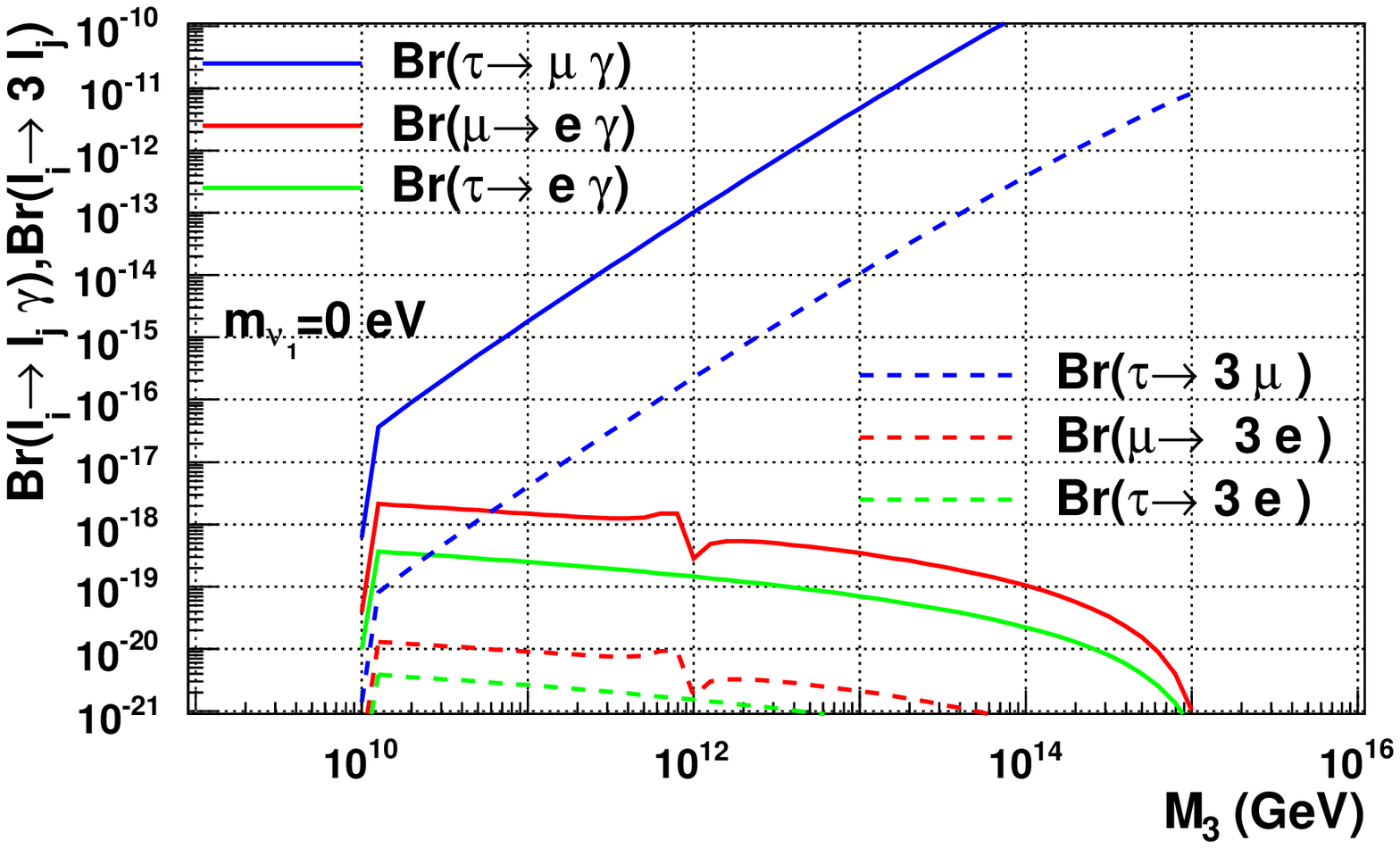}
\includegraphics[width=80mm,height=60mm]{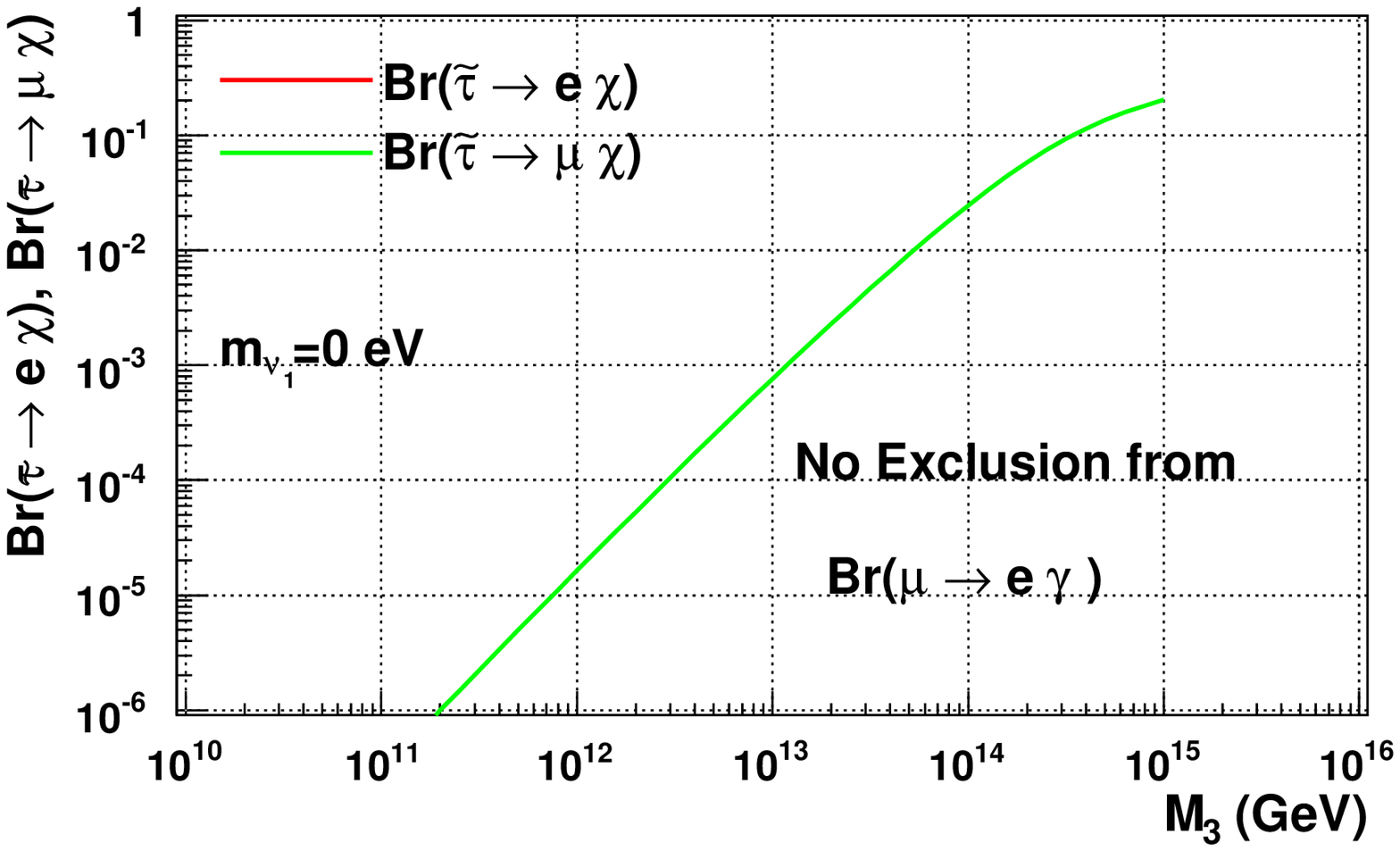}
\end{center}
\vspace{-5mm}
\caption{Branching ratios for $l_i \to  l_j + \gamma $  and 
$l_i \to 3 l_j$ (left) and LFV stau decays (right), for the standard 
point SPS3 as a function of $M_3$ for constant $M_1=M_2=10^{10}$ GeV.} 
\label{fig:LVF_M3}
\end{figure}

Finally, Fig.~\ref{fig:LVF_M3} shows branching ratios for $l_i \to
l_j + \gamma $ and $l_i \to 3 l_j$ (left) and LFV stau decays (right),
for the standard point SPS3 as a function of $M_3$ fixing
$s_{13}\equiv 0$ exactly. This implies that {\em all} final LFV states
involving electrons are tiny, as is expected from
Eq.~(\ref{eq:domM3}). Therefore for $s_{13}\equiv 0$ and $M_3$
dominance there is no constraint from the upper limit for Br$(\mu \to
e +\gamma)$. Once $s_{13}$ is nonzero branching ratios for LFV final
states involving electrons also become nonzero and proportional to
$s_{13}^2$.

In summary this section demonstrates that the full numerical
calculation confirms the analytical estimates presented above.
Absolute values of the LFV branching ratios for lepton decays are
sensitive functions of the unknown SUSY spectrum. For light sleptons,
usually the constraint from the non-observation of Br$(\mu \to e
+\gamma)$ makes the observation of LFV stau decays more likely when
$M_3$ gives the leading contribution to the LFV slepton mixing angles
and $s_{13}$ is close to zero. In this case LFV stau branching ratios
may exceed 10\%, as seen in Fig.~\ref{fig:LVF_M3}.
LFV stau branching ratios exceeding a percent are also possible for
SPS3 for hierarchical right-handed neutrinos and $M_1$ and $M_2$
dominance, as seen in Figs.~\ref{fig:StausLFV_M1} and
\ref{fig:LVF_M2}, but not for the SPS1a' case.
Similarly, for the case of degenerate neutrinos, LFV stau branching
ratios can exceed a few percent, as seen in Figs.~\ref{fig:BrSPS3},
especially for heavier sleptons, say $250-300$ GeV, where the Br$(\mu
\to e +\gamma)$ is smaller than the experimental limit and hence
does not place a restriction, as seen in Fig.~\ref{fig:Brs}.

Finally we note that we have expressed our results in terms 
of branching ratios. To get a rough idea on the observability 
of the signal, one has also to consider cross sections and backgrounds. 
For the signal itself one would have to work out a detailed set of cuts 
to suppress background which is clearly beyond the scope of the present 
work. However, after applying basic cuts used for SUSY signals 
\cite{DellaNegra:942733a} one can estimate the cross sections for 
$\tilde \tau_2$ production. Using {\tt PYTHIA 6.4} \cite{Sjostrand:2006za} 
we find for the sum of all (Drell-Yan) cross sections 126 fb (25 fb) and 
31 fb (3 fb) for $\tilde \tau_2$ in cascade decays in the case of SPS1a' 
(SPS3). Based on Monte Carlo analysis \cite{Hinchliffe:2001np,Hisano:2002tk} 
it has been shown that lepton flavour violation can be observed in dilepton 
invariant mass spectra within SUSY cascade decays. There the largest SM 
background is due to $t \bar{t}$ production. There is also SUSY background 
due to uncorrelated leptons stemming from different squark and gluino decay 
chains. The di-lepton spectra can provide a distinct signal of lepton 
flavour violation, namely the appearance of double peaks \cite{Bartl:2005yy} 
due to the fact that not only one but two or more sleptons can contribute 
to these spectra. In case of Drell-Yan processes the main background will 
be $W$ production.  To show more clearly the observability of such LFV 
signals a detailed Monte Carlo study would be necessary. This, 
however, is beyond the scope of the present paper.

We have shown results only for two ``standard'' mSugra points. 
However, as mentioned above, we have checked with a number of 
other points that {\em ratios of branching ratios} to a good 
approximation do not depend on the mSugra parameters. For 
absolute values of the branching ratios in general a heavier 
slepton spectrum leads to smaller LFV rates at low energy and 
larger LFV branching ratios at the LHC become possible, see also 
Fig.~\ref{fig:Brs}. Heavier sleptons, on the other hand, will 
lead to lower Drell-Yan production cross section, such that 
stau production will be dominated by cascade decays, the exact 
number of events depending on the details of the SUSY spectrum. 
We plan to do a more detailed, quantitative study of absolute 
event rates over all of mSugra space in the future.

\section{Conclusions and outlook}
\label{sec:cncl}

We have calculated lepton flavour violating processes both in LFV
decays of the $\mu$ and the $\tau$ leptons, as well as branching
ratios for LFV stau decays in the supersymmetric version of the
minimal type-I seesaw mechanism with mSugra boundary conditions.
We have limited ourselves to the study of a few standard mSugra 
points, ratios of LFV branching ratios are independent of this choice 
and therefore an interesting instrument to study the unknown seesaw 
parameters.

We have shown that the LFV branching ratios for lepton decays are 
sensitive functions of the unknown SUSY spectrum. For light sleptons, 
the non-observation of Br$(\mu \to e +\gamma)$ places an important 
constraint on the observability of LFV stau decays. The most favorable 
case is when right-handed neutrinos are hierarchical, with $M_3$ giving 
the leading contribution to the LFV slepton mixing angles and $s_{13}$ 
close to zero. In this case LFV stau branching ratios may exceed 10\% or 
so, see Fig.~\ref{fig:LVF_M3}. LFV stau branching ratios exceeding the 
percent level may also occur for hierarchical right-handed neutrinos 
with $M_1$ or $M_2$ dominance for the SPS3 reference point, but not for 
the SPS1a' case, see Figs.~\ref{fig:StausLFV_M1} and \ref{fig:LVF_M2}. 
Similarly, for the case of degenerate neutrinos, LFV stau branching 
ratios can exceed a few percent, as seen in Figs.~\ref{fig:BrSPS3}, 
especially for sleptons heavier than $250$ GeV or so, as seen in  
Fig.~\ref{fig:Brs}.

Notice that the above results rely crucially on an important
simplifying assumption about the right-handed neutrino spectrum.
For example, for degenerate right-handed neutrinos they require
that $R$ be real, while for hierarchical right-handed neutrinos
they hold when $R=1$.
This simplification allows one to calculate LFV decays of leptons and
of the scalar tau as a function of low-energy neutrino parameters.
However the use of this assumption should be critically scrutinized.
We plan to come back to this issue in a future publication.  Once an
improved experimental determination of $m_1$ and $s_{13}$ become
available from future double beta decay and neutrino oscillation
studies at reactor and accelerators, one could start ``learning''
about the right-handed neutrino sector, once the correct SUSY breaking
scheme has been identified and provided that the SUSY breaking scale
is above the lepton number breaking scale.

\section*{Acknowledgments}

Work supported by Spanish grants FPA2005-01269 and Accion Integrada
HA-2007-0090 (MEC) and by the European Commission network
MRTN-CT-2004-503369 and ILIAS/N6 RII3-CT-2004-506222.  The work of
A.V.M. is supported by {\it Funda\c c\~ao para a Ci\^encia e a
  Tecnologia} under the grant SFRH/BPD/30450/2006. W.P.~is partially
supported by the German Ministry of Education and Research (BMBF)
under contract 05HT6WWA, by the DAAD, project number D/07/13468 and 
by the 'Fonds zur F\"orderung der wissenschaftlichen Forschung' (FWF) 
of Austria, project. No. P18959-N16.

\end{document}